 \documentclass[lettersize,journal]{IEEEtran}
\IEEEoverridecommandlockouts
\usepackage{latexsym}
\usepackage{graphicx}
\usepackage{amsmath,amsfonts,amssymb}
\def\nba{{\boldsymbol{a}}}

\def\nbc{{\boldsymbol{c}}}
\def\nbd{{\boldsymbol{d}}}
\def\nbe{{\boldsymbol{e}}}

\def\nbg{{\boldsymbol{g}}}

\def\nbr{{\boldsymbol{r}}}
\def\nbs{{\boldsymbol{s}}}

\def\nbu{{\boldsymbol{u}}}
\def\nbv{{\boldsymbol{v}}}

\def\nbx{{\boldsymbol{x}}}

\def\nbz{{\boldsymbol{z}}}
\def\nb0{{\boldsymbol{0}}}
\def\nb1{{\boldsymbol{1}}}

\def\nbH{{\boldsymbol{H}}}
\def\nbI{{\boldsymbol{I}}}

\def\nbQ{{\boldsymbol{Q}}}
\def\nbR{{\boldsymbol{R}}}

\def\nbU{{\boldsymbol{U}}}
\def\nbV{{\boldsymbol{V}}}
\def\nbW{{\boldsymbol{W}}}

\def\ncalA{{\mathcal{A}}}

\def\Hhat{\widehat{H}}

\newtheorem{theorem}{Theorem}

\def\argmin{\operatorname{arg~min}}

\def\R{\mathbb{R}}

\def\N{\sigma^2}

\def\R{{\mathbb{R}}}

\usepackage[acronyms,nonumberlist,nopostdot,nomain,nogroupskip]{glossaries}
\newacronym{quic}{QUIC}{Quick UDP Internet Connections}
\newacronym{3gpp}{3GPP}{3rd Generation Partnership Project}
\newacronym{adc}{ADC}{Analog to Digital Converter}
\newacronym{5g}{5G}{5th generation}
\newacronym{aimd}{AIMD}{Additive Increase Multiplicative Decrease}
\newacronym{am}{AM}{Acknowledged Mode}
\newacronym{amc}{AMC}{Adaptive Modulation and Coding}
\newacronym{aqm}{AQM}{Active Queue Management}
\newacronym{awgn}{AGWN}{Additive White Gaussian Noise}
\newacronym{afd}{AFD}{Austin Fire Department}
\newacronym{balia}{BALIA}{Balanced Link Adaptation}
\newacronym{bdp}{BDP}{Bandwidth-Delay Product}
\newacronym{bf}{BF}{Beamforming}
\newacronym{cc}{CC}{Congestion Control}
\newacronym{cdf}{CDF}{Cumulative Distribution Function}
\newacronym{cn}{CN}{Core Network}
\newacronym{cqi}{CQI}{Channel Quality Information}
\newacronym{cp}{CP}{Control Plane}
\newacronym{csirs}{CSI-RS}{Channel State Information - Reference Signal}
\newacronym{dc}{DC}{Dual Connectivity}
\newacronym{dce}{DCE}{Direct Code Execution}
\newacronym{dci}{DCI}{Downlink Control Information}
\newacronym{dl}{DL}{Downlink}
\newacronym{dmr}{DMR}{Deadline Miss Ratio}
\newacronym{dmrs}{DMRS}{DeModulation Reference Signal}
\newacronym{e2e}{E2E}{End-to-End}
\newacronym{ecn}{ECN}{Explicit Congestion Notification}
\newacronym{edf}{EDF}{Earliest Deadline First}
\newacronym{enb}{eNB}{evolved Node Base}
\newacronym{epc}{EPC}{Evolved Packet Core}
\newacronym{es}{ES}{Edge Server}
\newacronym{fdma}{FDMA}{Frequency Division Multiple Access}
\newacronym{fdd}{FDD}{Frequency Division Duplexing}
\newacronym[firstplural=Radio Access Technologies (RATs)]{rat}{RAT}{Radio Access Technology}
\newacronym{fs}{FS}{Fast Switching}
\newacronym{ftp}{FTP}{File Transfer Protocol}
\newacronym{gnb}{gNB}{Next Generation Node Base}
\newacronym{harq}{HARQ}{Hybrid Automatic Repeat reQuest}
\newacronym{hetnet}{HetNet}{Heterogeneous Network}
\newacronym{hh}{HH}{Hard Handover}
\newacronym{hol}{HOL}{Head-of-Line}
\newacronym{ia}{IA}{Initial Access}
\newacronym{imt}{IMT}{International Mobile Telecommunication}
\newacronym{iot}{IoT}{Internet of Things}
\newacronym{los}{LOS}{Line of Sight}
\newacronym{lte}{LTE}{Long Term Evolution}
\newacronym{m2m}{M2M}{Machine to Machine}
\newacronym{mac}{MAC}{Medium Access Control}
\newacronym{mc}{MC}{Multi-Connectivity}
\newacronym{mcs}{MCS}{Modulation and Coding Scheme}
\newacronym{mec}{MEC}{Mobile Edge Cloud}
\newacronym{mi}{MI}{Mutual Information}
\newacronym{mimo}{MIMO}{Multiple Input, Multiple Output}
\newacronym{mmwave}{mmWave}{millimeter wave}
\newacronym{mr}{MR}{Maximum Rate}
\newacronym{mss}{MSS}{Maximum Segment Size}
\newacronym{mtd}{MTD}{Machine-Type Device}
\newacronym{mtu}{MTU}{Maximum Transmission Unit}
\newacronym{nfv}{NFV}{Network Function Virtualization}
\newacronym{nlos}{NLOS}{Non Line of Sight}
\newacronym{nr}{NR}{New Radio}
\newacronym{ofdm}{OFDM}{Orthogonal Frequency Division Multiplexing}
\newacronym{pdcch}{PDCCH}{Physical Downlonk Control Channel}
\newacronym{pdcp}{PDCP}{Packet Data Convergence Protocol}
\newacronym{pdsch}{PDSCH}{Physical Downlink Shared Channel}
\newacronym{pdu}{PDU}{Packet Data Unit}
\newacronym{pf}{PF}{Proportional Fair}
\newacronym{pgw}{PGW}{Packet Gateway}
\newacronym{phy}{PHY}{Physical}
\newacronym{pbch}{PBCH}{Physical Broadcast Channel}
\newacronym[plural=\gls{mme}s,firstplural=Mobility Management Entities (MMEs)]{mme}{MME}{Mobility Management Entity}
\newacronym{prb}{PRB}{Physical Resource Block}
\newacronym{pss}{PSS}{Primary Synchronization Signal}
\newacronym{pucch}{PUCCH}{Physical Uplink Control Channel}
\newacronym{pusch}{PUSCH}{Physical Uplink Shared Channel}
\newacronym{rach}{RACH}{Random Access Channel}
\newacronym{ran}{RAN}{Radio Access Network}
\newacronym{red}{RED}{Robotics Emergency Deployment}
\newacronym{rf}{RF}{Radio Frequency}
\newacronym{rlc}{RLC}{Radio Link Control}
\newacronym{rlf}{RLF}{Radio Link Failure}
\newacronym{rrc}{RRC}{Radio Resource Control}
\newacronym{rrm}{RRM}{Radio Resource Management}
\newacronym{rr}{RR}{Round Robin}
\newacronym{rs}{RS}{Remote Server}
\newacronym{rsrp}{RSRP}{Reference Signal Received Power}
\newacronym{rss}{RSS}{Received Signal Strength}
\newacronym{rtt}{RTT}{Round Trip Time}
\newacronym{rw}{RW}{Receive Window}
\newacronym{rx}{RX}{Receiver}
\newacronym{sa}{SA}{standalone}
\newacronym{sack}{SACK}{Selective Acknowledgment}
\newacronym{sap}{SAP}{Service Access Point}
\newacronym{sch}{SCH}{Secondary Cell Handover}
\newacronym{scoot}{SCOOT}{Split Cycle Offset Optimization Technique}
\newacronym{sdma}{SDMA}{Spatial Division Multiple Access}
\newacronym{sinr}{SINR}{Signal to Interference plus Noise Ratio}
\newacronym{sm}{SM}{Saturation Mode}
\newacronym{snr}{SNR}{Signal to Noise Ratio}
\newacronym{son}{SON}{Self-Organizing Network}
\newacronym{ss}{SS}{Synchronization Signal}
\newacronym{srs}{SRS}{Sounding Reference Signal}
\newacronym{sss}{SSS}{Secondary Synchronization Signal}
\newacronym{tb}{TB}{Transport Block}
\newacronym{tcp}{TCP}{Transmission Control Protocol}
\newacronym{tdd}{TDD}{Time Division Duplexing}
\newacronym{tdma}{TDMA}{Time Division Multiple Access}
\newacronym{tfl}{TfL}{Transport for London}
\newacronym{tm}{TM}{Transparent Mode}
\newacronym{trp}{TRP}{Transmitter Receiver Pair}
\newacronym{tti}{TTI}{Transmission Time Interval}
\newacronym{ttt}{TTT}{Time-to-Trigger}
\newacronym{tx}{TX}{Transmitter}
\newacronym{ue}{UE}{User Equipment}
\newacronym{ul}{UL}{Uplink}
\newacronym{uml}{UML}{Unified Modeling Language}
\newacronym{um}{UM}{Unacknowledged Mode}
\newacronym{utc}{UTC}{Urban Traffic Control}
\newacronym{vm}{VM}{Virtual Machine}
\newacronym{rsrq}{RSRQ}{Reference Signal Received Quality}
\newacronym{rssi}{RSSI}{Received Signal Strength Indicator}
\newacronym{crs}{CRS}{Cell Reference Signal}
\newacronym{comp}{CoMP}{Coordinated Multi-Point}
\newacronym{cran}{C-RAN}{Cloud \acrlong{ran}}
\newacronym{ca}{CA}{Carrier Aggregation}
\newacronym{cco}{CC}{Carrier Component}
\newacronym{nsa}{NSA}{Non Stand Alone}
\newacronym{embb}{eMBB}{Enhanced Mobility Broadband}
\newacronym{bsr}{BSR}{Buffer Status Report}
\newacronym{srb}{SRB}{Service Radio Bearer}
\newacronym{scm}{SCM}{Spatial Channel Model}
\newacronym{sctp}{SCTP}{Stream Control Transmission Protocol}
\newacronym{mptcp}{MPTCP}{Multi-path TCP}
\newacronym{ietf}{IETF}{Internet Engineering Task Force}
\newacronym{os}{OS}{Operating System}
\newacronym{tls}{TLS}{Transport Layer Security}
\newacronym{rfc}{RFC}{Request for Comments}
\newacronym{http}{HTTP}{HyperText Transfer Protocol}
\newacronym{nat}{NAT}{Network Address Translation}
\newacronym{api}{API}{Application Programming Interface}
\newacronym{rto}{RTO}{Retransmission Timeout}
\newacronym{psc}{PSC}{Public Safety Communication}
\newacronym{rpgm}{RPGM}{Reference Point Group Mobility}
\newacronym{ic}{IC}{Incident Command}
\newacronym{rsu}{RSU}{Road Side Unit}
\newacronym{uav}{UAV}{unmanned aerial vehicle}
\newacronym{usv}{USV}{Unmanned Surface Vehicle}
\newacronym{uas}{UAS}{Unmanned Aerial System}
\newacronym{iab}{IAB}{Integrated Access and Backhaul}
\newacronym{qoe}{QoE}{Quality of Experience}
\newacronym{ssim}{SSIM}{Structural Similarity Index}
\newacronym{psnr}{PSNR}{Peak Signal to Noise Ratio}
\newacronym{bs}{BS}{Base Station}
\newacronym{mu}{MU}{Multiple User}
\newacronym{ag}{AG}{Air-to-Ground}
\newacronym{af}{AF}{Array Factor}
\newacronym{ula}{ULA}{Uniform Linear Array}
\newacronym{upa}{UPA}{Uniform Planar Array}
\newacronym{lcs}{LCS}{Local Coordinate System}
\newacronym{psd}{PSD}{Power Spectral Density}
\newacronym{vq}{VQ}{vector quantization}
\newacronym{a2g}{A2G}{air-to-ground}
\newacronym{em}{EM}{electromagnetic}
\newacronym{vae}{VAE}{variational autoencoder}

\def\bb0{{\mathbb{0}}}

\def\bb{{\boldsymbol{b}}}

\def\b0{{\boldsymbol{0}}}

\def\b{{\mathrm{b}}}

\def\r0{{\mathbf{0}}}

\def\bsf0{{\bm{\mathsf{0}}}}

\def\N0{{N_{\mathrm{0}}}}

\def\bsf{{\boldsymbol{s}_\mathrm{f}}}

\newcommand{\be}{\begin{equation}}
\newcommand{\ee}{\end{equation}}
\newcommand{\bal}{\begin{align}}
\newcommand{\eal}{\end{align}}

\usepackage{textcomp}
\usepackage{gensymb}
\usepackage{siunitx}
\usepackage{xcolor}
\usepackage{tikz}
\usepackage{etoolbox}
\usepackage{mathtools}
\usepackage{subcaption}
\usetikzlibrary{chains,arrows,calc,positioning}
\usepackage{multirow}
\usepackage{comment}
\usepackage{cite}

\usepackage{booktabs}

\def\BibTeX{{\rm B\kern-.05em{\sc i\kern-.025em b}\kern-.08em T\kern-.1667em\lower.7ex\hbox{E}\kern-.125emX}}

\newtoggle{conf}
\toggletrue{conf}

\newtoggle{conference}
\togglefalse{conference}

\newtoggle{onecolumn}
\togglefalse{onecolumn}  

\iftoggle{onecolumn}{
    \usepackage[font=small]{caption}
    \usepackage{sidecap}
}
{}

\begin{document}
%
\title{Parametrization and Estimation of High-Rank Line-of-Sight MIMO Channels  with Reflected Paths}

\author{
%
\IEEEauthorblockN{Yaqi Hu, \IEEEmembership{Student Member, IEEE}, Mingsheng Yin, \IEEEmembership{Student Member, IEEE},\\
Sundeep Rangan, \IEEEmembership{Fellow, IEEE}, Marco Mezzavilla, \IEEEmembership{Senior Member, IEEE}} \\

\thanks{The authors were supported 
by NSF grants 
1952180, 1925079, 1564142, 1547332, the SRC, OPPO,
and the industrial affiliates of NYU WIRELESS. The work was also supported by Remcom that provided the Wireless Insite  
software.}
}

\maketitle

\begin{abstract}
High-rank line-of-sight (LOS) MIMO systems
have attracted considerable attention for
millimeter wave and THz communications.
The small wavelengths in these frequencies
enable spatial multiplexing with massive data rates at long distances.
Such systems are also being considered for 
multi-path non-LOS (NLOS) environments. 
In these scenarios, standard channel
models based on plane waves cannot capture the
curvature of each wave front necessary to model
spatial multiplexing.
This work presents a novel and
simple multi-path wireless channel parametrization 
where each path is replaced by a LOS path with a 
reflected image source.  
The model is fully valid for all paths
with specular planar reflections,
and captures 
the spherical nature of each wave front.
Importantly, it is shown that the model  uses
only two additional parameters relative to the standard
plane wave model.  Moreover, the parameters
can be easily captured in standard ray tracing.  The 
accuracy of the approach is demonstrated
on detailed ray tracing simulations 
at \SI{28}{GHz} and \SI{140}{GHz} in a dense
urban area.
\end{abstract}


\section{Introduction}
\emph{Line-of-sight (LOS) multi-input multi-output (MIMO)} systems \cite{bohagen2005construction,  sarris2007design, matthaiou2008capacity,wei2021channel}
have emerged as a valuable technology
for the millimeter wave (mmWave) and terahertz (THz) frequencies.  The concept
is to operate communication links
at a transmitter-receiver (TX-RX) separation,
$R$, less than the so-called Rayleigh distance,
\begin{equation} \label{eq:rayleigh}
  R < R_{\rm rayleigh} \approx \frac{2D^2}{\lambda},
\end{equation}
where $D$ is the total aperture of the 
TX and RX arrays and $\lambda$ is the wavelength.
In this regime, links can support
multiple spatial streams even with a single
LOS path~\cite{winters1987capacity}.
LOS MIMO is particularly valuable
in the mmWave and THz frequencies, where the
wavelength $\lambda$ is small and hence
the Rayleigh distance --- which sets the
maximum range of such systems --- can be large
with moderate size apertures $D$.
For example, at \SI{140}{GHz} carrier
frequency with an aperture of $D=$\,\SI{1}{m},
the Rayleigh distance is $R_{\rm rayleigh} \approx $\,\SI{930}{m} enabling long range communication
while remaining below the Rayleigh distance.

Indeed, there have been several demonstrations
in the \SI{60}{GHz} mmWave bands
\cite{sheldon200860ghz}. Also, with the advancement of  communication systems in the THz and sub-THz bands \cite{kurner2014towards},
there has been growing interest in high-rank LOS MIMO in higher frequencies as well \cite{singh2019challenges}
-- see, for example, some recent work at \SI{140}{GHz} \cite{zaman2018140,sawaby2020fully}. 


Many applications for such LOS MIMO systems are envisioned as operating in NLOS settings.
For example, in mid-haul and 
backhaul applications -- a key target
application for sub-THz systems \cite{sawaby2020fully,gougeon2020assessment,chintareddy2021preliminary} -- 
NLOS paths may be present from ground clutter
when serving street-level radio units.
In this work, we will use the term \emph{wide aperture MIMO},
instead of \emph{LOS MIMO}, since we are also interested
in cases where the systems operate in such
NLOS settings.

Evaluating wide aperture systems in NLOS environments requires accurate channel models
to describe multi-path propagation.
Conventional statistical multipath
models, such as  QuaDRiGA \cite{jaeckel2014quadriga} and 3GPP \cite{3GPP38901},
describe each path as a propagating
plane wave with a gain, delay, and directions of arrival and departure.
Under this standard plane wave approximation
(PWA), the MIMO channel response can be computed
for any array geometries at the TX and RX
\cite{heath2018foundations}.
However, the PWA model is not valid when the TX-RX separation is below the Rayleigh distance (i.e., not in the 
far-field), since the
curvature of each wavefront becomes important.
While spherical wave models are well-understood
for single LOS path channels 
\cite{bohagen2009spherical}, 
there are currently few techniques to model
them in NLOS multi-path settings.

\iftoggle{onecolumn}{  
    \begin{SCfigure}[][t!]
    \includegraphics[width=0.35\textwidth]{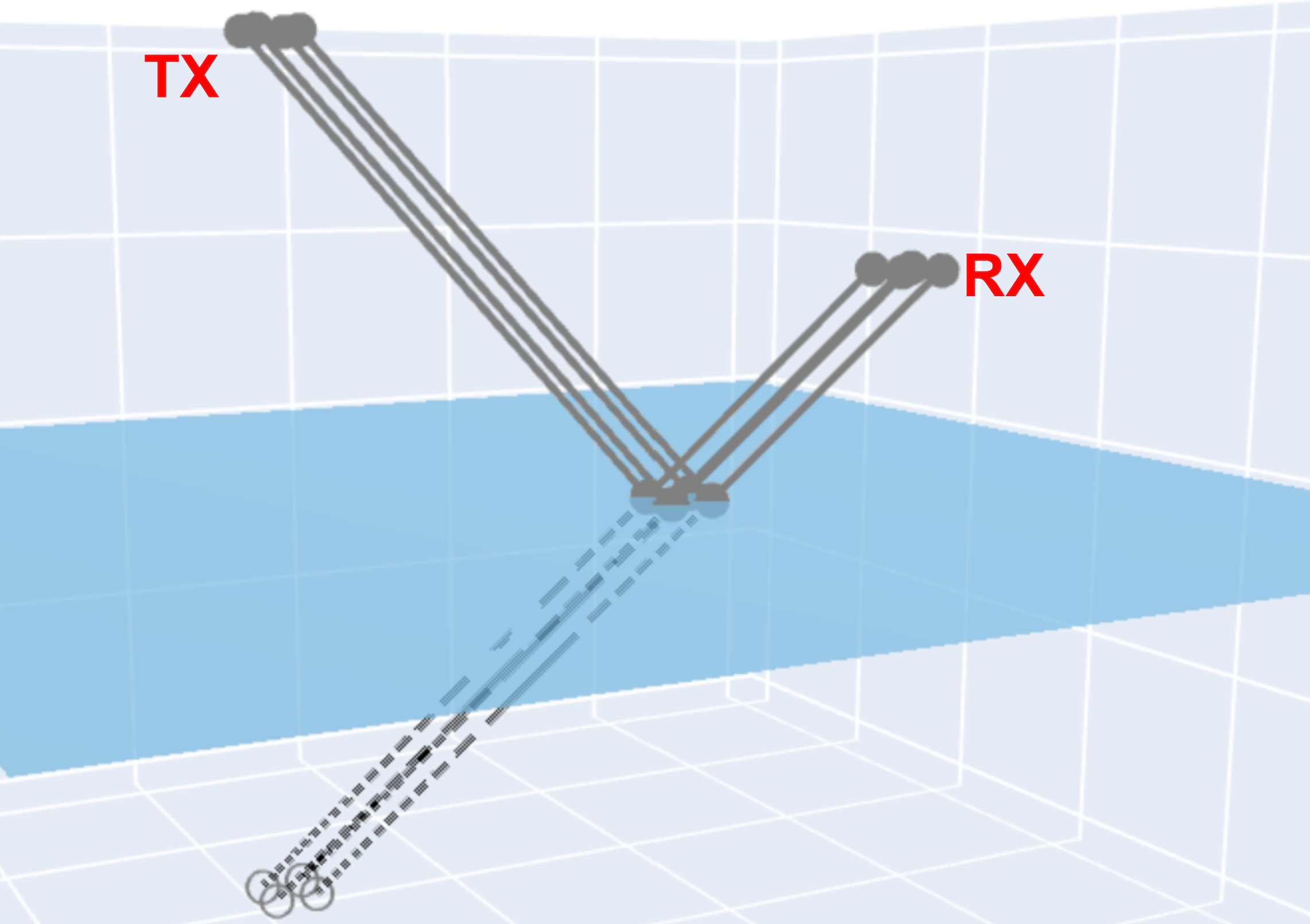}
    \caption{The proposed reflection model
    replaces each reflected path with a LOS path
    with a mirror image source.  The spherical 
    wave from this source can then be easily modeled
    for arbitrarily wide aperture arrays.}
    \label{fig:mirror_rx}
    \vspace{-5mm}
    \end{SCfigure} 
}{
    \begin{figure}
    \centering
    \includegraphics[width=0.8 \linewidth]{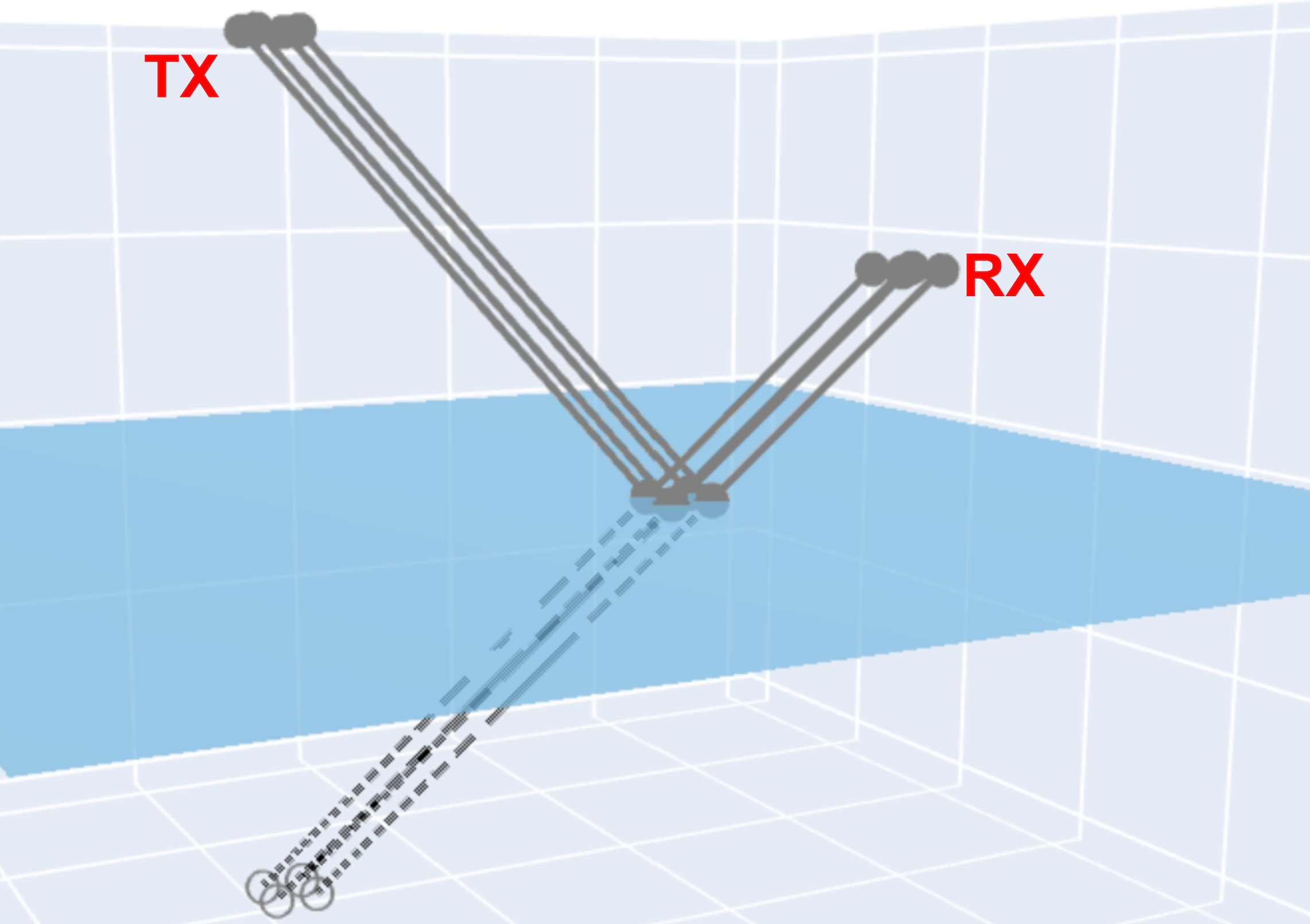}
    \caption{The proposed reflection model
    replaces each reflected path with a LOS path
    with a mirror image source.  The spherical 
    wave from this source can then be easily modeled
    for arbitrarily wide aperture arrays.}
    \label{fig:mirror_rx}
    \end{figure}
}

In this work, we present a simple parametrization
for multipath channels that capture the full
spherical nature of each wavefront.  The model
is valid for both LOS paths as well as 
NLOS paths arising from arbitrary numbers of
specular reflections from flat surfaces 
(i.e., no curvature).  
The main concept is that,
in such environments, each NLOS path
can be replaced by a LOS path where the
TX location is replaced by a virtual 
image source from the reflection on the 
source -- See Fig.~\ref{fig:mirror_rx}.
This idea is the same concept that underlies the
method of images in ray tracing \cite{yun2015ray}.
See, also \cite{pizzo2022line}
for modeling reflections in the near field.
Our main contribution here 
shows that the propagation from 
each such image can be parametrized
by two additional parameters relative to the plane
wave model.  We call the parametrization the
\emph{reflection model}, or RM.

In addition, we show how these parameters can be
extracted for site-specific evaluations
via ray tracing.  
Analyzing wide aperture systems with 
ray tracing 
typically requires running the simulations between each transmitter and receiver element pair,
which can be computationally expensive when
the number of elements is large.
In addition, the ray tracing must be repeated for different antenna geometries or orientations, 
making site planning and capacity evaluation
time-consuming.  
In contrast, we show how
the full parameters for the RM can be computed from 
ray tracing a \emph{single} ray tracing
simulation near the array centers.
As an illustration, we show an example application of using the RM model 
to estimate
the MIMO capacity in a point-to-point link
with dramatically lower ray tracing simulation
time than would be required exhaustive 
ray tracing.  
\paragraph*{Prior Work}  As stated above,
most current industry models, such as 
\cite{jaeckel2014quadriga} and \cite{3GPP38901},
use a plane wave approximation, which 
is only valid in the far-field.  Obtaining
the exact near-field behavior, generally 
requires performing ray tracing between 
each TX and RX element.  In this sequel, 
we will call this method \emph{exhaustive}
ray tracing.  While accurate, exhaustive
ray tracing is computationally expensive.  
The closest related line of work to finding
computationally simpler models can be found
in the recent papers \cite{cui2022channel, yu2022hybrid, cui2023near, wei2021channel}.  These works consider
near field channels from point \emph{scatterers}
close to the receiver.  In contrast, the present paper considers reflections from \emph{surfaces} in 
the near-field of the transmitter or receiver.  
A key difference with surfaces is that the point
of reflection is different for different
TX and RX element positions.

\section{Plane Wave Approximations for Multi-Path Channels}
We begin by reviewing the standard
plane wave multi-path channel models
using the perspective in \cite{bohagen2009spherical}.
Consider a wireless channel from a
TX locations  $\nbx^t \in \ncalA^t$
to RX locations $\nbx^r \in \ncalA^r$,
where $\ncalA^t$ and  $\ncalA^r \subset \R^p$
are some regions that can contain
the elements in the TX and RX arrays.
We focus on 
so-called 3D models with $p=3$,
although similar results can be derived
for $p=2$.
We assume the channel is described by a set of $L$ discrete 
paths representing the routes of propagation from
the TX to RX locations.
In this case, the channel frequency response
at a frequency $f$ from a transmit
location $\nbx^t$ to a receive location
$\nbx^r$ is given by:

\begin{equation}  \label{eq:Hf}
     H(f) = \sum_{\ell=1}^L 
    g_\ell \exp\left(-\frac{j2\pi  f}{c} d_\ell(\nbx^r,\nbx^t) \right),
\end{equation}

where, for each path $\ell=1,\ldots,L$, 
$g_\ell$ is a complex nominal channel gain 
(assumed to be approximately constant over the region), $d_\ell(\nbx^r,\nbx^t)$ is the propagation distance
along the path from $\nbx^t$ to $\nbx^r$,
and $c$ is the speed of light
\cite{heath2018foundations}.
We will call $d_\ell(\nbx^r,\nbx^t)$ the \emph{path
 distance function} for  path $\ell$.

Describing the gain $g_\ell$ and path distance function $d_\ell(\cdot)$
for each path is sufficient to compute the response
for arbitrary TX and RX arrays in multi-path
environments.  For example,
suppose that the TX array has $N_{\rm tx}$
elements at locations $x^t_n  \in \ncalA^t$,
$n=1,\ldots,N_{\rm tx}$ and the RX array
has elements at locations $x^r_m \in \ncalA^r$,
$m=1,\ldots,N_{\rm rx}$.
Then, the MIMO frequency response
is the matrix with coefficients
\begin{equation}  \label{eq:Hmn}
    H_{mn}(f) = \sum_{\ell=1}^L 
    g_\ell \exp\left(-\frac{j2\pi f}{c} d_\ell(\nbx^r_m,\nbx^t_n) \right).
\end{equation}
Hence, if we can find the gain $g_\ell$ and path distance function $d_\ell(\nbx^r,\nbx^t)$ for each
path, we can compute the wideband MIMO channel response.

The main challenge is how to model the
path distance function $d_\ell(\nbx^r,\nbx^t)$
as a function of the RX and TX positions $\nbx^r$ and $\nbx^t$.
If path $\ell$ is LOS,
the path distance function 
is simply the Euclidean distance
\begin{equation}     \label{eq:dlos}
    d_\ell(\nbx^r,\nbx^t) = \|\nbx^r-\nbx^t\|.
\end{equation}
For NLOS paths, the distance function is usually 
approximated under the assumption that
the propagation in each path are plane waves.
Specifically, suppose that 
$\nbx^r_0$ and $\nbx^t_0$ are some reference locations
for the RX and TX.  For example, these points
could be the
centroids of the arrays.
Now, for small displacements 
$\nbx^t - \nbx^t_0$
and $\nbx^r - \nbx^r_0$, one often 
assumes a plane wave approximation (PWA)

\iftoggle{onecolumn}{ 
    \begin{align}  \label{eq:dpwa}
        \MoveEqLeft 
        d_\ell(\nbx^r,\nbx^t) \approx 
        \widehat{d}_\ell(\nbx^r,\nbx^t) = c \tau_\ell 
        + (\nbu^r_\ell)^\intercal(\nbx^r_0 - \nbx^r) 
        + (\nbu^t_\ell)^\intercal(\nbx^t_0 - \nbx^t),
    \end{align}   
}{
   \begin{align}  \label{eq:dpwa}
        \MoveEqLeft 
        d_\ell(\nbx^r,\nbx^t) \approx 
        \widehat{d}_\ell(\nbx^r,\nbx^t) \nonumber \\
        &= c \tau_\ell 
        + (\nbu^r_\ell)^\intercal(\nbx^r_0 - \nbx^r) 
        + (\nbu^t_\ell)^\intercal(\nbx^t_0 - \nbx^t),
    \end{align}
}

where $c$ is the speed of light,
$\tau_\ell$ is the time of flight between the nominal points $\nbx^r_0$ and $\nbx^t_0$
along the path,
\begin{equation}
    c\tau_\ell = d_\ell(\nbx^r_0,\nbx^t_0),
\end{equation}
and $\nbu^r_\ell$ and $\nbu^t_\ell$ are unit vectors in $\R^p$ representing the directions of arrival
and departure of the path.  
We will call \eqref{eq:dpwa}
the \emph{PWA model}.

When path $\ell$ is a LOS path, so that
$d_\ell(\nbx^r,\nbx^t)$ is given by 
\eqref{eq:dlos}, the parameters for the 
PWA model \eqref{eq:dpwa} are
\begin{subequations} \label{eq:pwa_los}
\begin{align}
    \tau_\ell &= \frac{1}{c} \|\nbx^r_0-\nbx^t_0\| \\
    \nbu^r &= \frac{\nbx^t_0-\nbx^r_0}{c\tau}, \quad
    \nbu^t = \frac{\nbx^r_0-\nbx^t_0}{c\tau}.
\end{align}
\end{subequations}

The direction vectors $\nbu^r_\ell$ and $\nbu^t_\ell$
are also the negative derivatives of the distance
function at $(\nbx^r,\nbx^t)=(\nbx^r_0,\nbx^t_0)$, meaning:

\begin{equation} \label{eq:uderiv}
    (\nbu^r)^\intercal = -\frac{\partial d_\ell(\nbx^r_0,\nbx^t_0)}{\partial \nbx^r},
    \quad
    (\nbu^t)^\intercal= -\frac{\partial d_\ell(\nbx^r_0,\nbx^t_0)}{\partial \nbx^t}.
\end{equation}

Hence, the PWA model is valid to a second-order error
approximation in that

\iftoggle{onecolumn}{  
   \begin{align} 
         \MoveEqLeft d_\ell(\nbx^r,\nbx^t) -
        \widehat{d}_\ell(\nbx^r,\nbx^t) 
        = O(\|\nbx^r_0 - \nbx^r\|^2)
        + O(\|\nbx^t_0 - \nbx^t\|^2).    \label{eq:err_pwa}
    \end{align} 
}{
   \begin{align} 
         \MoveEqLeft d_\ell(\nbx^r,\nbx^t) -
        \widehat{d}_\ell(\nbx^r,\nbx^t) \nonumber \\
        &= O(\|\nbx^r_0 - \nbx^r\|^2)
        + O(\|\nbx^t_0 - \nbx^t\|^2).    \label{eq:err_pwa}
    \end{align}
}

Typically, we express the directions $\nbu^r_\ell$ and $\nbu^t_\ell$
in spherical coordinates.  For $p=3$, we can write
these unit vectors as
\begin{subequations} \label{eq:usph}
\begin{align}
    \nbu^r_\ell &= (\cos(\phi^r_\ell)\cos(\theta^r_\ell), 
    \sin(\phi^r_\ell)\cos(\theta^r_\ell), \sin(\theta^r_\ell)) \\
    \nbu^t_\ell &= (\cos(\phi^t_\ell)\cos(\theta^t_\ell), 
    \sin(\phi^t_\ell)\cos(\theta^t_\ell), \sin(\theta^t_\ell)),
\end{align}
\end{subequations}
where $\phi^r_\ell, \phi^t_\ell$ are the azimuth
AoA and AoD for path $\ell$,
and $\theta^r_\ell, \theta^t_\ell$ are
the elevation AoA and AoD. 
Thus, the channel can be described by six parameters
per path with a total of $6L$ parameters:
\begin{equation} \label{eq:param_pwa}
    (g_\ell, \tau_\ell, \phi^r_\ell,
    \theta^r_\ell,\phi^t_\ell,\theta^t_\ell),
    \quad
    \ell=1,\ldots,L.
\end{equation}

The PWA model \eqref{eq:dpwa}
thus has clear benefits:
it is geometrically interpretable and accurate when
the total array aperture is small.  The main disadvantage is that
it becomes inaccurate when the array aperture is large and higher-order terms of the displacements
$\nbx^r - \nbx^r_0$ and $\nbx^t - \nbx^t_0$ become
significant.  In particular, the PWA
model always predicts that each path contributes
at most one spatial rank.  
But, for wide aperture
arrays, the spherical nature of the wavefront
can result in a higher rank channel even for a single path \cite{bohagen2005construction,  sarris2007design, matthaiou2008capacity,wei2021channel}.
In particular, a channel with only a LOS path
can have a higher spatial rank, but the PWA model
will not be able to predict this feature.

In contrast, the path distance function 
\eqref{eq:dlos} is exact for arbitrary displacements.
However, this model is only valid for LOS paths.
The question is whether there is a model
for the path distance function that is exact 
for arbitrary array sizes and applies in NLOS
settings.

\section{Modeling the Distance Function under Planar, Specular Reflections}

\subsection{The Reflection Model}

Our first result provides a 
geometric characterization 
of the path distance function for paths
with arbitrary numbers of specular reflections
from flat planes.
Specifically, we show that the path distance of the reflected path 
is identical to a LOS distance to a rotated and translated image point.
Moreover, the parameters for the rotation and reflection can be derived from
the path route.  The result does not apply to 
 curved surfaces, diffractions, or scattering.
However, we will show in the simulations below
that, even in a realistic environment with these properties, as well as 
losses such as foliage, the model
performs well.   

To state the result, let $\ncalA^t$ and $\ncalA^r
\subset \R^p$ be regions of space.  Suppose that
for every TX location $\nbx^t \in \ncalA^t$
and RX location $\nbx^r \in \ncalA^r$ there is a path that has a constant
set of reflecting surfaces where each surface is a plane.  
In this case, we will say the path \emph{has 
constant planar reflections} over the regions
$\ncalA^t$ and $\ncalA^r$.  With this definition,
our first result is as follows:

\medskip
\begin{theorem} \label{thm:path_dist_orthogonal}
Suppose a path has constant planar 
reflections from regions
$\ncalA^t$ to $\ncalA^r \subset \R^p$.  
Let $\nbx^t_0 \in \ncalA^t$
and $\nbx^r_0 \in \ncalA^r$ be arbitrary points
in these regions.  Then, 
there exists an orthogonal matrix $\nbU \in \R^{p \times p}$
and vector $\nbg \in \R^p$ such that
\begin{equation} \label{eq:dnlosUg}
    d(\nbx^r,\nbx^t) = \left\| \nbx^r - \nbU\nbx^t
    -\nbg \right\|.
\end{equation}
\end{theorem}
\begin{IEEEproof}
\iftoggle{onecolumn}{  
   \begin{SCfigure}[][t!]
      \includegraphics[width=0.32\linewidth]{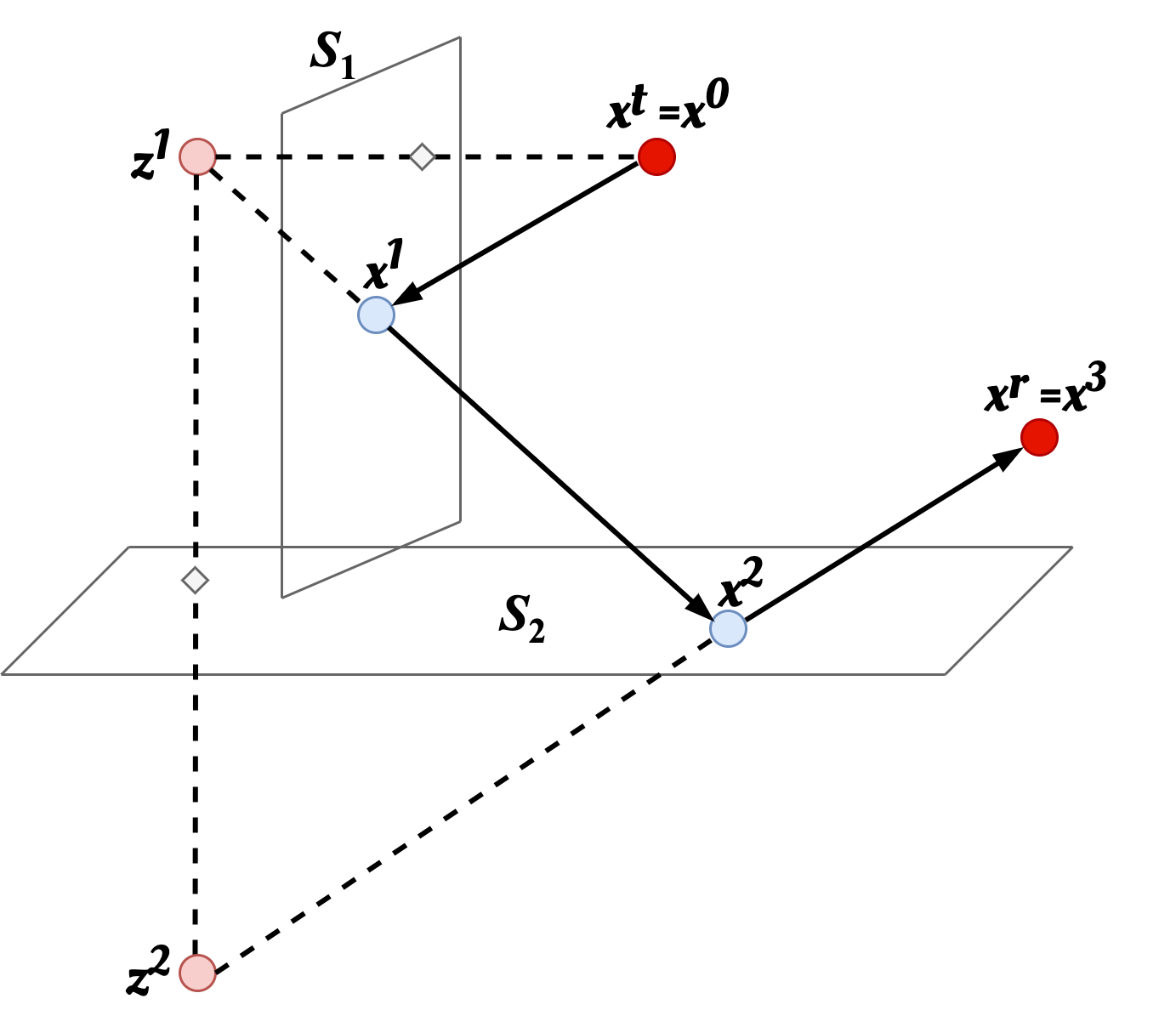}
      \caption{Example route with two reflections (i.e, $K=3$). The initial point is 
      the transmitter, $\nbx^t = \nbx^0$.  The final point is the receiver,
      $\nbx^r = \nbx^K$. The intermediate points, $\nbx^1, \nbx^2$ are the locations
      of the reflections on the surfaces denoted $S_1$ and $S_2$.}
      \label{fig:2timeRef}
      \vspace{-4mm}
    \end{SCfigure}
}{
   \begin{figure}
      \centering
      \includegraphics[width=0.8\linewidth]{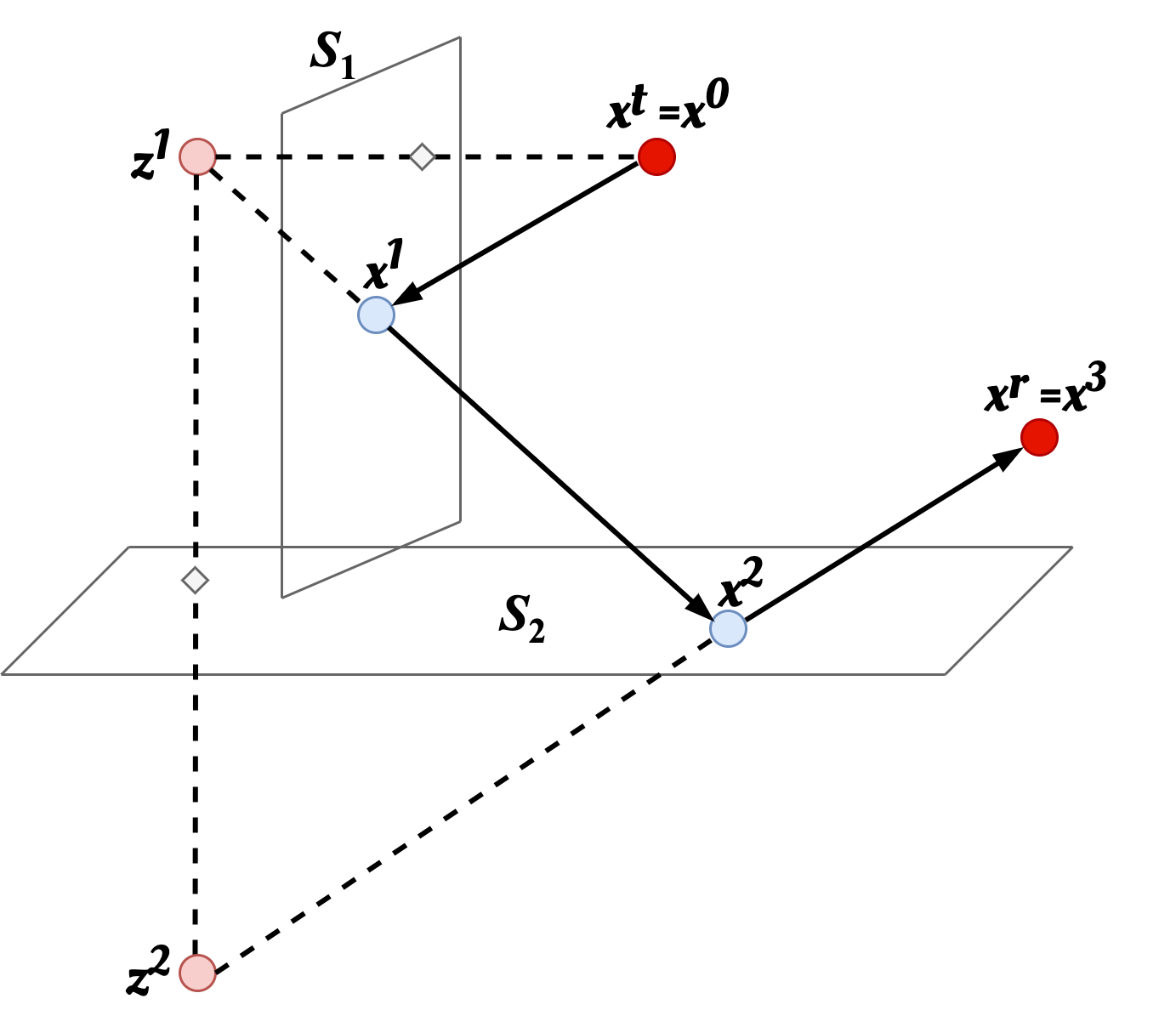}
      \caption{Example route with two reflections (i.e, $K=3$). The initial point is 
      the transmitter, $\nbx^t = \nbx^0$.  The final point is the receiver,
      $\nbx^r = \nbx^K$. The intermediate points, $\nbx^1, \nbx^2$ are the locations
      of the reflections on the surfaces denoted $S_1$ and $S_2$.}
      \label{fig:2timeRef}
    \end{figure}
}

Write the path's route 
as a sequence of $K-1$ interactions:
\begin{equation} \label{eq:path_sequence}
    \nbx^t = \nbx^0 \rightarrow \nbx^1 \rightarrow \cdots
    \rightarrow \nbx^K = \nbx^r,
\end{equation}
where, $\nbx^i$ represents the location of the $i$-the reflection.
The initial point, $\nbx^0 = \nbx^t$, is the TX location
and the final point, $\nbx^K = \nbx^r$, is the RX location.
An example path with two reflections (i.e., $K=3$) 
is shown in Fig.~\ref{fig:2timeRef}.
Let $S_k$ denote the $k$-th reflecting plane.
For example, in Fig.~\ref{fig:2timeRef},
there are two surfaces, $S_1$ and $S_2$.

The key idea in the proof is to trace the 
reflection of the transmitted point in each surface.
Simple geometry shows that, after $K$ reflections,
the image of the transmitted point is an orthogonal
and shifted transformation of the original point,
and the path distance is the distance to this reflected point.
The details are in Appendix~\ref{sec:proof_theorem1}.
\end{IEEEproof}

\medskip 
We note that the distance function
\eqref{eq:dnlosUg} has a simple geometric interpretation:
The distance of any reflected path
is identical to the distance on a LOS path
but with the TX in a rotated
and shifted the reference frame.
The rotation is represented by the 
orthogonal matrix $\nbU$ and the
shift by the vector $\nbg$.
Returning to Fig.~\ref{fig:mirror_rx},
this rotated and shifted point is simply the mirror
image of the original transmitter.

It is important to recognize that, 
in a multi-path channel, there will be 
separate parameters, $\nbU_\ell$ and $\nbg_\ell$,
for each path $\ell$.  Thus, if we are computing a 
MIMO channel matrix component, $H_{mn}(f)$
in \eqref{eq:Hmn}, each path distance function 
must be computed from an expression of the form:
\[
    d_\ell(\nbx^r_m,\nbx^t_n) = \left\| \nbx^r_m - \nbU_\ell \nbx^t_n
    -\nbg_\ell \right\|.
\]
We will discuss how to estimate the parameters
$(\nbU_\ell,g_\ell)$ for each path
in Section~\ref{sec:param_fit}.

\subsection{Relation to the PWA Model}

The description \eqref{eq:dnlosUg} can also be 
easily connected to the parameters in the PWA.
Let $R_z(\phi)$, $R_y(\theta)$
and $R_x(\gamma)$ be the rotation matrices
around the $z$, $y$ and $x$ axes:

\begin{subequations} \label{eq:rotmatrix}
\begin{align} 
    \nbR_z(\phi) &:= \begin{bmatrix}
        \cos(\phi) & -\sin(\phi) & 0 \\
        \sin(\phi) & \cos(\phi) & 0 \\
        0 & 0 & 1 
        \end{bmatrix}, \\ 
    \nbR_y(\theta) &:= \begin{bmatrix}
        \cos(\theta) & 0 & \sin(\theta) \\
        0 & 1 & 0 \\
        -\sin(\theta) & 0 & \cos(\theta) 
        \end{bmatrix} \\
    \nbR_x(\gamma) &:= 
     \begin{bmatrix}
        1 & 0 & 0 \\
        0 & \cos(\gamma) & -\sin(\gamma) \\
        0 & \sin(\gamma) &  \cos(\gamma) 
        \end{bmatrix} 
\end{align}
\end{subequations}
Also, for $s=\pm 1$, let 
$\nbQ_z(s)$ be the reflection in the $z$-axis:
\begin{equation} \label{eq:Qzdef}
    \nbQ_z(s) := \begin{bmatrix}
        1 & 0 & 0 \\
        0 & 1 & 0 \\
        0 & 0 & s
        \end{bmatrix}.
\end{equation}
With these definitions, we have the following result.

\begin{theorem} \label{thm:path_dist_param}
Suppose a path has $K-1$ 
constant planar reflections from  regions
$\ncalA^t$ to $\ncalA^r \subset \R^p$.  
Let $\nbx^t_0 \in \ncalA^t$
and $\nbx^r_0 \in \ncalA^r$ be arbitrary points
in these regions.  Then, if $p=3$, there exists parameters
\begin{equation} 
    (\tau,\phi^r,\theta^r,\phi^t,\theta^t,\gamma^t,s)
\end{equation}
where 
\begin{equation} 
    \label{eq:sdef}
    s = 
    \begin{cases}
        -1 & \mbox{if $K$ is even} \\
        1 & \mbox{if $K$ is odd},
    \end{cases}
\end{equation}
such that for all $\nbx^r \in \ncalA^r$ and $\nbx^t \in \ncalA^t$, the total path distance is
\iftoggle{onecolumn}{  
   \begin{align} 
        \MoveEqLeft
            d(\nbx^r,\nbx^t) = \Bigl\| 
            c\tau\nbe_x +  \nbR_y(\theta^r)\nbR_z(-\phi^r)(\nbx^r_0-\nbx^r)
             + \nbQ_z(s)\nbR_x(\gamma^t)
           \nbR_y(\theta^t)\nbR_z(-\phi^t)(\nbx^t_0-\nbx^t) 
        \Bigr\|.
        \label{eq:dnlos}
    \end{align}
}{
   \begin{align} 
        \MoveEqLeft
            d(\nbx^r,\nbx^t) = \Bigl\| 
            c\tau\nbe_x +  \nbR_y(\theta^r)\nbR_z(-\phi^r)(\nbx^r_0-\nbx^r)
            \nonumber \\
           & + \nbQ_z(s)\nbR_x(\gamma^t)
           \nbR_y(\theta^t)\nbR_z(-\phi^t)(\nbx^t_0-\nbx^t) 
        \Bigr\|.
        \label{eq:dnlos}
    \end{align}
}
Moreover the parameters $(\tau,\phi^r,\theta^r,\phi^t,\theta^t)$
match the parameters in the PWA model
\eqref{eq:dpwa}.
\end{theorem}
\begin{IEEEproof}
See Appendices~\ref{sec:proof_path_dist_param} and
Appendix~\ref{sec:proof_equiv}.
\end{IEEEproof}

\medskip
The importance
of the result is that, in $p=3$,
the path distance function can be explicitly
written as a set of rotation angles:
Specifically, there are 
elevation and azimuth angles,
$\theta^r$ and $\phi^r$ at the RX,
and roll, elevation, and azimuth angles
    $\gamma^t$, $\theta^t$ and $\phi^t$
at the TX.  There is an additional
binary reflection term $s$.

In a multi-path channel, there will
be one set of such parameters
for each path along with a path gain.  
Thus, if there are $L$ paths, the 
parameters for the channel will be
\begin{equation} \label{eq:param_high}
    (g_\ell, \tau_\ell, \phi^r_\ell,
    \theta^r_\ell,\phi^t_\ell,\theta^t_\ell,\gamma^t_\ell,s_\ell), \quad 
    \ell=1,\ldots,L
\end{equation}
where we have added the complex gain $g_\ell$ 
and delay $\tau_\ell$ for each path $\ell$.
In comparison to the PWA model \eqref{eq:param_pwa}, 
there is one additional binary parameter $s_\ell = \pm 1$ and one additional angle $\gamma^t_\ell$
per path.
We have thus found a concise parametrization
of the distance function that is exact and valid
for all paths with arbitrary planar reflections.
We will call the parametrization \eqref{eq:param_high} the \emph{reflection model} (RM).

\section{Fitting the RM 
Parameters from Ray Tracing Data} \label{sec:param_fit}
A benefit of the PWA model is that
computing the terms of MIMO channel matrix
\eqref{eq:Hmn} is computationally simple.
Specifically, one typically only needs
to run ray tracing once 
between any reference TX and 
RX locations $\nbx_0^t$ and $\nbx^r_0$
near the array elements.  If the PWA parameters for the paths
\eqref{eq:param_pwa} can be extracted from those simulations, then
for any elements close to the reference locations,
the path distance and phase offset of 
the path can be computed from \eqref{eq:dpwa}.

Unfortunately, this approach is not valid for wide 
aperture arrays where the displacements
$\nbx^t-\nbx^t_0$ and $\nbx^r-\nbx^r_0$
are large.  In ray tracing, the conventional approach is to 
perform a simulation for each pair of TX and RX elements to
capture the full MIMO response accurately.
Hence, if there are $N_{\rm tx}$ and $N_{\rm rx}$ elements on the 
TX and RX arrays, the computational complexity
grows by $N_{\rm rx}N_{\rm tx}$.  Moreover, if the arrays are
moved or changed, the simulations need to be performed again.  As ray tracing is computationally costly,
performing ray tracing $N_{\rm rx}N_{\rm tx}$ times
for each possible array configuration or orientation 
can be computationally prohibitive.

In contrast, if one has the full 
RM model parameters \eqref{eq:param_high}
for each path, the MIMO channel matrix
coefficients \eqref{eq:Hmn} can 
be computed for arbitrary array 
geometries without re-running the ray tracing.

In this section, we show how the model parameters 
\eqref{eq:param_high}
can be extrapolated
from a limited number of ray tracing simulations.  
We describe two potential methods:
\begin{itemize}
    \item \emph{RM via Route Tracing (RM-RT)}:  In this method, we assume that we can obtain
    the full route \eqref{eq:path_sequence}
    for each path.  This route is provided by most ray tracers,
    such as the commercial Wireless Insite
    ray tracer \cite{Remcom} that we use below.  
    Given the route information, we show that the complete set of 
    RM model parameters \eqref{eq:param_high} can be found directly from 
    the channel from a single pair of reference
    locations $(\nbx^t_0,\nbx^r_0)$.
    
    \item \emph{RM via Displaced Pairs (RM-DP)}:  In this case,
    we assume the ray tracing provides  only the PWA parameters 
    \eqref{eq:param_pwa} for any TX-RX pair.  However,
    the path route \eqref{eq:path_sequence} is not provided.
    In this case, we show that the RM model 
    parameters can be found from the PWA model parameters at $M+1$
    TX-RX pairs with one pair being a reference pair, and $M$ additional pairs at locations displaced from the reference.  The number of required additional pairs 
    is $M \geq 2$.
\end{itemize}

\subsection{Parameter Estimation via Route Tracing}
\label{sec:routetracing}
In the first method, \emph{reflection model via route tracing (RM-RT)}, ray tracing 
is performed between some reference TX and RX pair locations, $\nbx_0^t$ and $\nbx_0^r$.
We assume that, in addition to the  PWA parameters \eqref{eq:param_pwa},
the ray tracing provides the physical 
\emph{route} of each multi-path component.  
Specifically, for each path, we assume the ray tracer provides 
sequence of points as in \eqref{eq:path_sequence}.
Under this assumption, we can obtain the parameters $\nbU$ and $\nbg$ in 
\eqref{eq:dnlosUg} following the proof of Theorem~\ref{thm:path_dist_orthogonal}.
The steps are as follows:
\begin{enumerate}
\item Compute the direction vector, $\nbv^k$, 
of each step from \eqref{eq:dir_vec}.
\item Compute the normal vector, $\nbu^k$, and 
intercept, $b^k$, 
to the $k$-th interacting surface from \eqref{eq:norm_vec} and \eqref{eq:intercept}.
\item Compute the reflection matrix $\nbV_k$ and translation
vector $\nbc^k$ in \eqref{eq:Vkck}.
\item Compute the sequence of intercepts, $\nbg^k$, from \eqref{eq:grec}.
\item Compute $\nbU$ and $\nbg$ from \eqref{eq:Uprod}.
\end{enumerate}
After finding $\nbU$ and $\nbg$, 
we can also find the parameters \eqref{eq:param_high}
in Theorem~\ref{thm:path_dist_param} from the proof
of that theorem:
\begin{enumerate}
    \item Compute, $\nbd_0$, 
    the separation vector from the RX
    to the reflection TX image from \eqref{eq:dreflect}.
    \item Compute the angles $(\theta^r.\phi^r)$
    and distance $\tau$ by putting $\nbd_0$ into
    spherical coordinates \eqref{eq:dspherical}.
    \item Compute the binary term $s=\pm 1$
    from the number of reflections, $K-1$, using
    \eqref{eq:sdef}.
    \item Compute $\nbW$ from \eqref{eq:Wdef}.
    \item Since $\nbQ_z(s)\nbW$ is an orthogonal matrix
    with determinant one, write the matrix
    as a product of rotation matrices \eqref{eq:VprodR1}
    to recover the TX angles $(\gamma^t,\theta^t,\phi^t)$ \cite{lynch2017modern}.
\end{enumerate}
Again, note that this procedure is performed on
each path.  Hence, if the ray tracing provides $L$
paths, the procedure will be performed $L$ times,
producing parameters \eqref{eq:param_high}
for $\ell=1,\ldots,L$.

\subsection{Parameter Estimation via Displaced Pairs}
\label{sec:rmdp}
In this case, we assume the ray tracer does not include
full path route
\eqref{eq:path_sequence} between TX-RX pairs.
Instead, the ray tracer only provides the standard
PWA parameters \eqref{eq:param_pwa} for each TX-RX pair.
To obtain the RM parameters, we will perform 
ray tracing between
a total of $M+1$ TX-RX pairs 
$(\nbx^t_m,\nbx^r_m)$, $m=0,1,\ldots,M$.
By convention, we will call the first pair, $(\nbx^t_0,\nbx^r_0)$,
the \emph{reference pair} and the remaining $M$
pairs $(\nbx^t_m,\nbx^r_m)$, $m=1,\ldots,M$,
the \emph{displaced pairs}.  
Between each TX-RX pair, $m=0,\ldots,M$, we assume 
we have PWA parameters of the form:
\begin{equation} \label{eq:meas_rmdp}
    (g_{\ell m},\tau_{\ell m}, \phi_{\ell m}^r,\theta_{\ell m}^r, 
        \phi_{\ell m}^t,\theta_{\ell m}^t), \quad
        \ell = 1,\ldots, L_m,
\end{equation}
where $L_m$ is the number of paths in pair $m$,
$(\phi_{\ell m}^r,\theta_{\ell m}^r, 
        \phi_{\ell m}^t,\theta_{\ell m}^t)$
are the angles of arrival and departure of
path $\ell$, $g_{\ell m}$ is its complex gain,
and $\tau_{\ell m}$ is its absolute delay.
We show in Appendix~\ref{sec:displaced_details}
that if we have this from $M \geq 2$ TX-RX displaced pairs, 
we can solve for all the parameters  \eqref{eq:param_high} in the RM model.
We will call the RM parameters estimated 
from this procedure RM via Displaced Pairs or RM-DP.

Under the ideal assumptions of the theory
-- namely that all reflections are specular from surfaces
with no curvature -- the RM-RT and RM-DP methods
will return the same parameters for the RM model.
However, most ray tracers also model other interactions
including diffractions, diffuse reflections, and transmissions.  In addition, the curvature of surfaces may also be accounted for.  In this case, RM-RT and RM-DP
may return slightly different results.
However, we will see in the simulations below
that the differences are small.



\iftoggle{onecolumn}{  
   \begin{SCfigure}[][t]
      \centering
      \includegraphics[width=0.6\linewidth]{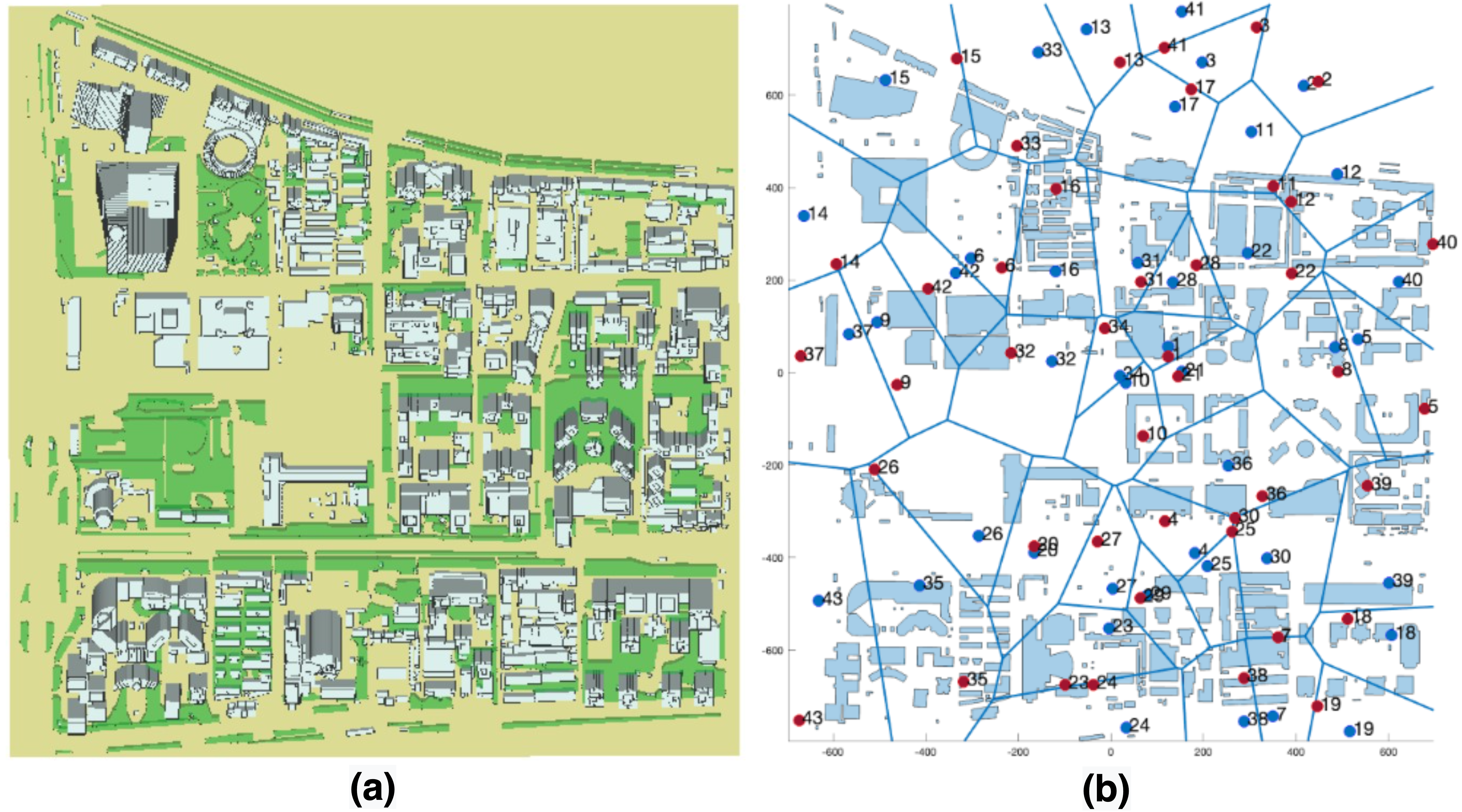}
      \caption{(a) An example of the ray tracing environment: a section of Beijing city. Foliage areas are indicated by the green blocks. (b) The distribution of reference TX and RX pairs. The transmitters and receivers are dropped randomly by implementing a Voronoi partition within our partial Beijing area.}
      \label{fig:remcom_beijing}
      \vspace{-3mm}
    \end{SCfigure}
}{
   \begin{figure}
      \centering
      \includegraphics[width=1\linewidth]{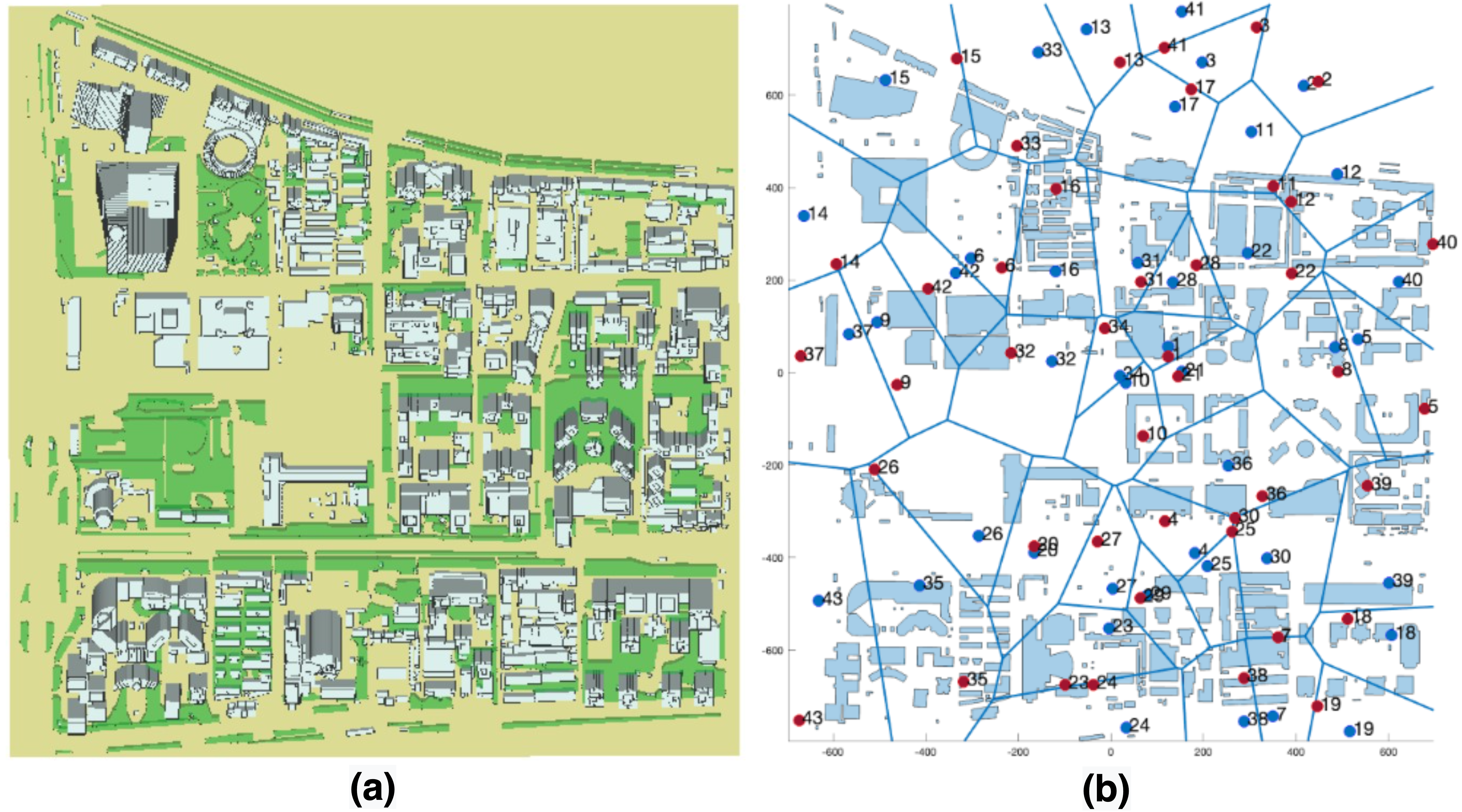}
      \caption{(a) An example of the ray tracing environment: a section of Beijing city. Foliage areas are indicated by the green blocks. (b) The distribution of reference TX and RX pairs. The transmitters and receivers are dropped randomly by implementing a Voronoi partition within our partial Beijing area.}
      \label{fig:remcom_beijing}
    \end{figure}
}

\section{Validation in an Urban Environment}
\label{sec:valid}

The RM model is exact under the ideal assumption that 
paths remain constant over the region of interest
and that all reflections are specular and planar.
Of course, in reality, these assumptions may not be exactly valid
and hence the RM model may still have some errors,
particularly when we are trying to estimate the channel
at displacements far from the reference.
To quantify this error, we conducted a ray tracing simulation
of a 1650 $\times$ 1440 square meter
area of a 
dense urban environment of Beijing, China, as 
shown in Fig.~\ref{fig:remcom_beijing}. The identical ray
 tracing environment was used 
 in the channel modeling work
 \cite{XiaRanMez2020}.
Within this area, we selected $N=43$
TX and RX pairs spaced within \SI{200}{m} of each other.
We call each of these pairs the \emph{reference locations}.
For each such reference pair $(\nbx^r_0,\nbx^t_0)$,
we also generated $M=6$ random \emph{displaced}
locations  $(\nbx^r_m,\nbx^t_m)$, $m=1,\ldots,M$
with distances from \SI{1}{cm} to \SI{100}{cm}
from the reference location.  The displaced pairs are shown
in Fig.~\ref{fig:move_direction}.
We then run complete ray tracing between each pair
 $(\nbx^r_m,\nbx^t_m)$, $m=0,\ldots,M$,
to obtain PWA parameters \eqref{eq:meas_rmdp}.
The ray tracing is performed at two different reference RF frequencies $f_0=$\,\SI{28}{GHz} and \SI{140}{GHz}.

Importantly, 
the links in the test scenario
have a significant fraction of energy
where the RM model may not be exact.
Table~\ref{tb:power_contri} shows the average
percentage of power contributions of the 
ray-traced paths at the \SI{100}{cm} displacement
for four categories:  line of sight (LOS), reflections only, foliage, and diffraction without foliage.  The RM model is theoretically exact only
for the LOS and reflection paths, which constitute
less than 50\% of the energy for at both \SI{28}{GHz} and \SI{140}{GHz}.  Nevertheless,
we will see that the RM model is able
to predict the channel well.

\begin{table}[]
\centering
\caption {Power percentage contribution} 
\label{tb:power_contri} 
\begin{tabular}{|l|c|c|}
\hline
Power percentage                                                                 & 28 GHz-100cm & 140 GHz-100cm \\ \hline
Paths with LoS                                                                   & 38.36        & 48.28         \\ \hline
Paths with reflection only                                                       & 4.35         & 4.00          \\ \hline
Paths with foliage                                                               & 57.28        & 47.71         \\ \hline
\begin{tabular}[c]{@{}l@{}}Paths with diffraction \\ and no foliage\end{tabular} & 0.01         & 0.01          \\ \hline
\end{tabular}
\end{table}

\iftoggle{onecolumn}{  
   \begin{SCfigure} 
      \centering
      \includegraphics[width=0.35 \linewidth]{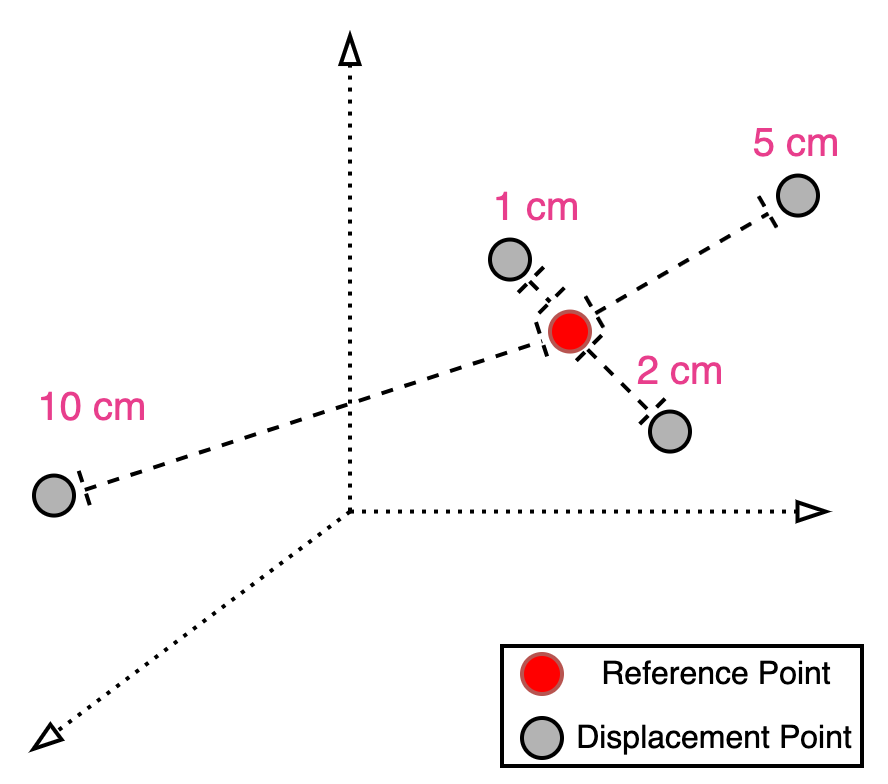}
      \caption{Random displaced pairs generation for testing:
      Around each location $(\nbx^r_0,\nbx^t_0)$, we generated displaced locations at random 3D locations
      $(\nbx^r_m,\nbx^t_m)$ at distances: \SI{1}{cm}, \SI{2}{cm}, \SI{5}{cm},
      \SI{10}{cm}, \SI{50}{cm} and \SI{100}{cm}.}
      \vspace{-5mm}
      \label{fig:move_direction}
    \end{SCfigure}
}{
   \begin{figure} 
      \centering
      \includegraphics[width=0.6 \linewidth]{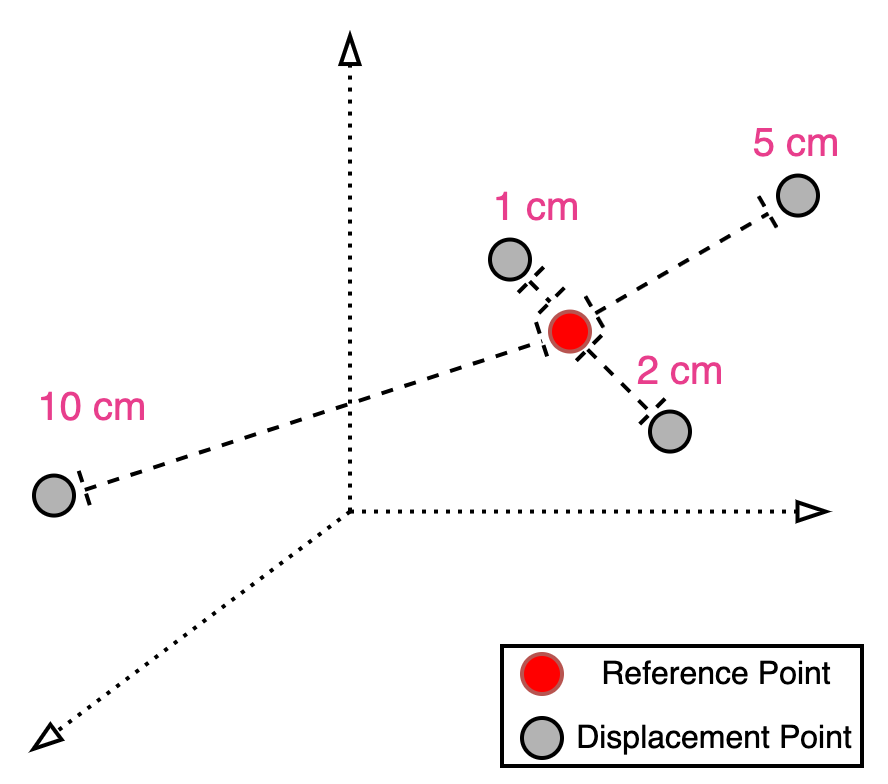}
      \caption{Random displaced pairs generation for testing:
      Around each location $(\nbx^r_0,\nbx^t_0)$, we generated displaced locations at random 3D locations
      $(\nbx^r_m,\nbx^t_m)$ at distances: \SI{1}{cm}, \SI{2}{cm}, \SI{5}{cm},
      \SI{10}{cm}, \SI{50}{cm} and \SI{100}{cm}.}
      \vspace{-5mm}
      \label{fig:move_direction}
    \end{figure}
}
 
Now, let $H_m(f)$ denote the complex 
channel from the $\nbx^t_m$ to $\nbx^r_m$ 
at an RF frequency $f$.  
The ``true" value of this channel
can be computed from the ray tracing data between the pair
$(\nbx^r_m,\nbx^t_m)$ at the reference RF frequency $f_0$.
Specifically, the complex channel gain at
any RF frequency $f$ is given by
\begin{equation} \label{eq:Htrue}
        H_m(f) = 
        \sum_{\ell=1}^{L_m} g_{\ell m} e^{-j2\pi  (f-f_0)\tau_{\ell m}},
\end{equation}
where $L_m$ is the number of paths between 
$\nbx^t_i$ and $\nbx^r_i$; and for path $\ell$, 
$g_{\ell m}$ is the complex gain of the path
at the reference location and frequency,
and $\tau_{\ell m}$ is its delay.

We wish to see how well different models 
can predict the true channels
$H_m(f)$
between the displaced TX-RX pairs $(\nbx^r_m,\nbx^t_m)$,
using ray tracing information only near 
the reference TX-RX pair $(\nbx^r_0,\nbx^t_0)$.
We compare three methods:
\begin{itemize}
    \item \emph{Constant model}:  
    $\Hhat_m(f) = H_0(f)$ where we assume that
    the channel does not change from the
    reference location.
    
    \item \emph{PWA model:}  The estimate
     is computed from
    \begin{align}    
    \label{eq:Hhat}
      \Hhat_m(f) = 
        \sum_{\ell=1}^{L_0} g_{\ell 0} \exp\left[j2\pi 
        \left(\tau_{\ell 0}f_0 -
         \frac{f\widehat{d}_{\ell m}}{c}         \right) \right],
     \end{align}     
    where $\widehat{d}_{\ell m}$ is the estimate of the distance
    $d_{\ell}(\nbx^r_m,\nbx^t_m)$ on path $\ell$ 
    from the PWA model \eqref{eq:dpwa}:
\iftoggle{onecolumn}{  
   \begin{align}     \label{eq:dpwa_test}
        \MoveEqLeft \widehat{d}_{\ell m} = \widehat{d}_{\ell}(\nbx^r_m,\nbx^t_m) = c \tau_{\ell 0} 
        + (\nbu^r_{\ell 0})^\intercal(\nbx^r_0 - \nbx^r_m) 
        + (\nbu^t_{\ell 0})^\intercal(\nbx^t_0 - \nbx^t_m),
    \end{align}
}{
   \begin{align}     \label{eq:dpwa_test}
        \MoveEqLeft \widehat{d}_{\ell m} = \widehat{d}_{\ell}(\nbx^r_m,\nbx^t_m) = c \tau_{\ell 0} 
        \nonumber \\
        &+ (\nbu^r_{\ell 0})^\intercal(\nbx^r_0 - \nbx^r_m) 
        + (\nbu^t_{\ell 0})^\intercal(\nbx^t_0 - \nbx^t_m),
    \end{align}
}
where $\tau_{\ell 0}$ is the delay between the reference pair
$(\nbx^r_0,\nbx^t_0)$ and $\nbu^r_{\ell 0}$ and $\nbr^t_{\ell 0}$
are the unit vectors in the directions of arrival and departure
at the reference location:
\begin{subequations} \label{eq:usph_ref}
\begin{align}
    \nbu^r_{\ell 0} &= (\cos(\phi^r_{\ell 0})\cos(\theta^r_{\ell 0}), 
    \sin(\phi^r_{\ell 0})\cos(\theta^r_{\ell 0}), \sin(\theta^r_{\ell 0})) \\
    \nbu^t_{\ell 0} &= (\cos(\phi^t_{\ell 0})\cos(\theta^t_{\ell 0}), 
    \sin(\phi^t_{\ell 0})\cos(\theta^t_{\ell 0}), \sin(\theta^t_{\ell 0})).
\end{align}
\end{subequations}
The channel estimate \eqref{eq:Hhat} thus represents the estimate
based on extrapolated path distances using the PWA parameters
from the reference.
    
    \item \emph{Reflection model (RM)}:  For the reflection model,
    we compute the RM model parameters $(\nbU_\ell,\nbg_\ell)$ for
    all paths $\ell$ using either the RM-RT or RM-DP methods
    in Section~\ref{sec:param_fit}.  We then use 
    channel estimate
     \eqref{eq:Hhat} where 
     $\widehat{d}_{\ell m}$ are the estimates of the distances
    $d(\nbx^r_m,\nbx^t_m)$ computed 
    from the reflection model \eqref{eq:dnlosUg}:
\begin{equation} \label{eq:drm_test}
\widehat{d}_{\ell m} = \widehat{d}_{\ell}(\nbx^r_m,\nbx^t_m) = 
\left\| \nbx^r_m - \nbU_\ell\nbx^t_m
    -\nbg_\ell \right\|.
\end{equation}
Equivalently, we can obtain the RM parameters \eqref{eq:param_high}
and use the distance \eqref{eq:dnlos}.  These two parametrizations
will give the same answer.
\end{itemize}

For the reflection model, the parameters were extracted
as described in Section~\ref{sec:param_fit} by 
implementing both the RM-RT and RM-DP methods.
In the RM-RT method, the high-precision coordinates of all interaction points between reference TX-RX pairs are exported from the ray tracer.
For the RM-DP method, we used $M=2$ for the two displaced pairs at the distances  \SI{1}{cm} and \SI{2}{cm} from the reference location. 
We used $M=2$ since, as discussed
above, this value is the minimum number to 
uniquely identify the parameters.
These are the two displaced points closest to the reference.
 
\iftoggle{onecolumn}{  
   \begin{figure*}
    \footnotesize
        \centering
        \begin{subfigure}[b]{0.7\textwidth}
            \centering
            \includegraphics[width=\textwidth]{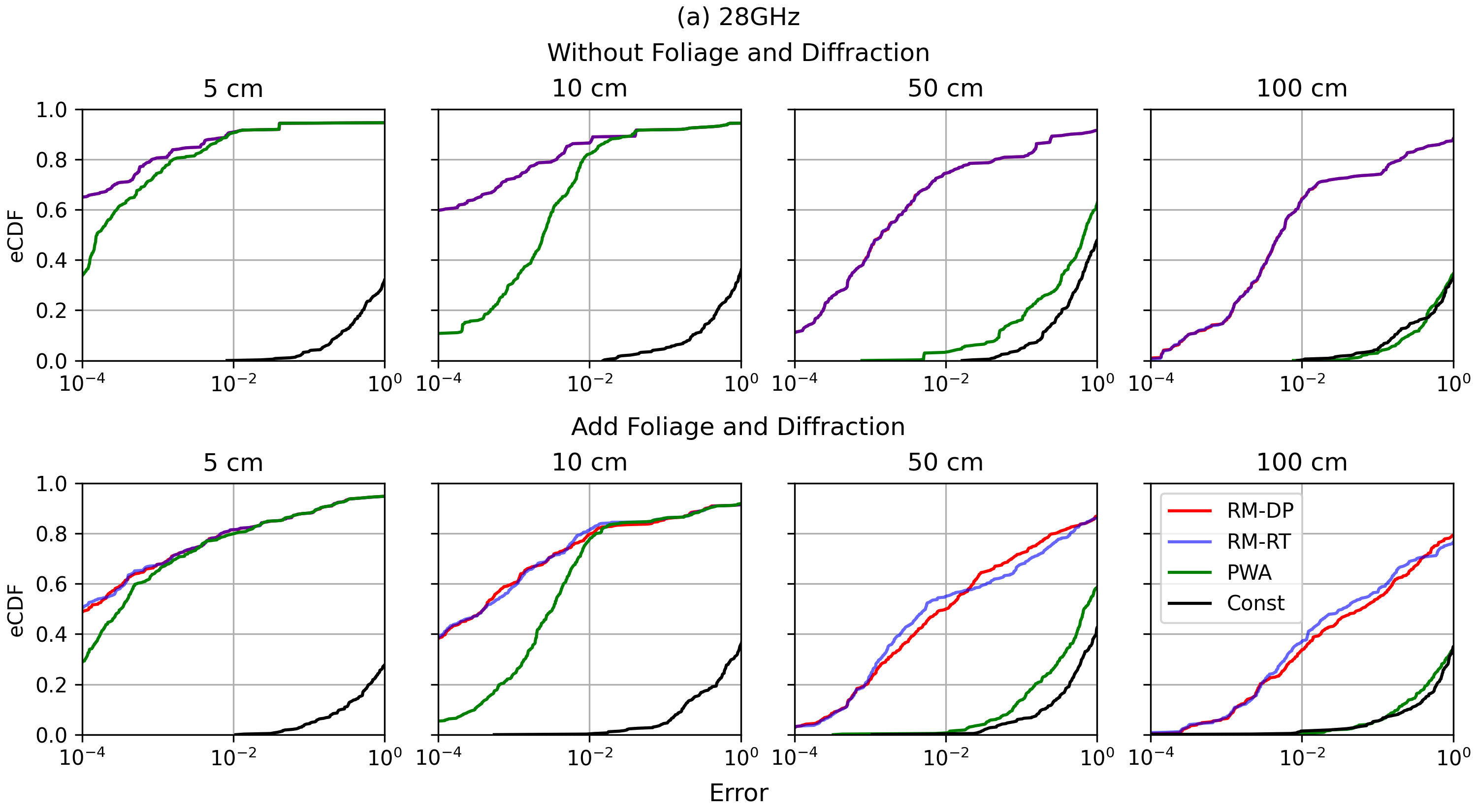}
        \end{subfigure}
        \vskip\baselineskip
        \begin{subfigure}[b]{0.7\textwidth}
            \centering
            \includegraphics[width=\textwidth]{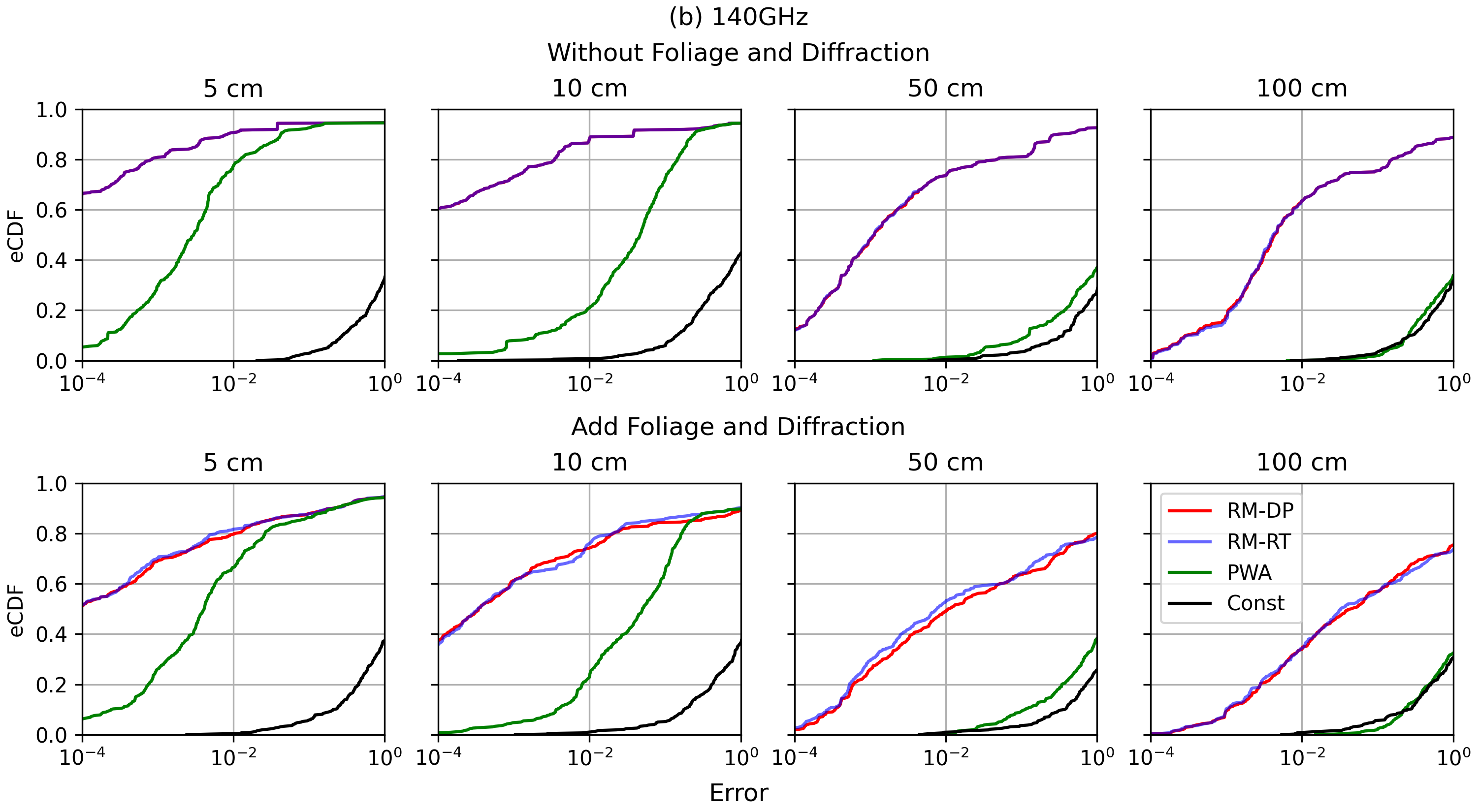}
        \end{subfigure}
        \caption{eCDF plot for the error of estimated channel gain in the randomized directions at different distances. The displacement distances are set to \SI{5}{cm}, \SI{10}{cm}, \SI{50}{cm}, and \SI{100}{cm}, where (a) is \SI{28}{GHz}; (b) is \SI{140}{GHz}.
        The RM-DP refers to the estimation of reflection model parameters via displaced pairs, while the RM-RT refers to the estimation of reflection model parameters via route tracing.
        And the PWA refers to the standard plane wave approximation model.}
    \label{fig:ecdf_lambda_rnd_28_140}
    \vspace{-2mm}
    \end{figure*}
}{
   \begin{figure*}
    \footnotesize
        \centering
        \begin{subfigure}[b]{0.85\textwidth}
            \centering
            \includegraphics[width=\textwidth]{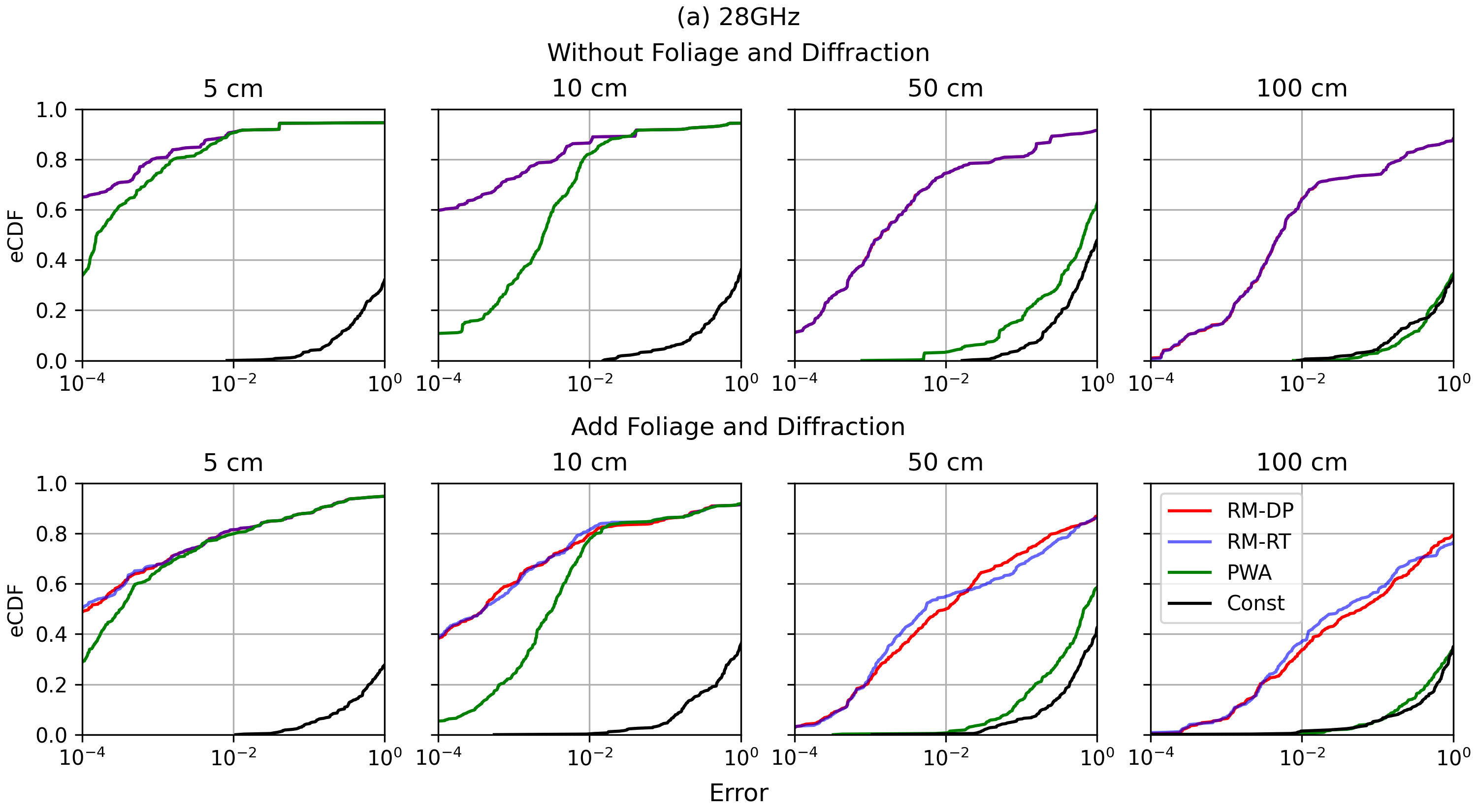}
        \end{subfigure}
        \vskip\baselineskip
        \begin{subfigure}[b]{0.85\textwidth}
            \centering
            \includegraphics[width=\textwidth]{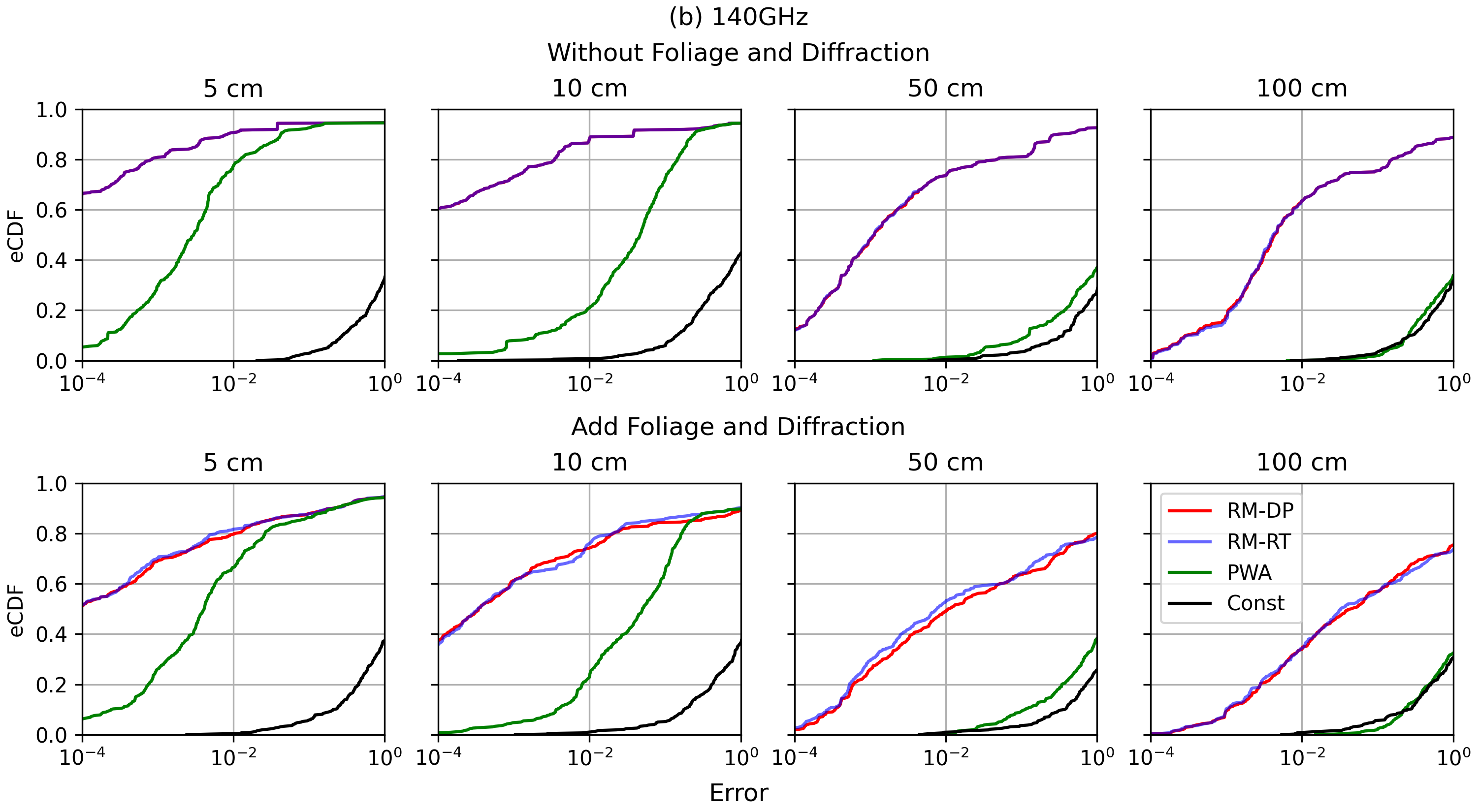}
        \end{subfigure}
        \caption{eCDF plot for the error of estimated channel gain in the randomized directions at different distances. The displacement distances are set to \SI{5}{cm}, \SI{10}{cm}, \SI{50}{cm}, and \SI{100}{cm}, where (a) is \SI{28}{GHz}; (b) is \SI{140}{GHz}.
        The RM-DP refers to the estimation of reflection model parameters via displaced pairs, while the RM-RT refers to the estimation of reflection model parameters via route tracing.
        And the PWA refers to the standard plane wave approximation model.}
    \label{fig:ecdf_lambda_rnd_28_140}
    \vspace{-2mm}
    \end{figure*}
}

Similar to \cite{skrimponis2020power}, 
we performed the validation on two bands:
\SI{28}{GHz} with a bandwidth of \SI{400}{MHz},
and \SI{140}{GHz} with a bandwidth of \SI{2}{GHz}. 
On each link, the true and estimated channels were computed
at the reference and displaced locations at ten
random frequencies within the bandwidth.  
All ray tracing was performed using Wireless Insite
by Remcom \cite{Remcom}.
Importantly, the modeling also includes diffraction, so that
deviations from the theory due to non-specular reflections are included.
Additionally, the ray tracing can be run with or without foliage.
Since interactions with foilage do not necessarily follow the theory,
this feature will also enable us to measure the accuracy of the model
under more realistic propagation mechanisms.
The source code and data for the validation process can be found at \cite{wide-aperture}.


We compute the normalized mean squared errors:
\begin{equation} \label{eq:ref_err}
    \epsilon_{m}(f)  := \frac{\left|\Hhat_m(f) - H_m(f) \right|^{2}}{E_0},
    \quad
    E_0 := \sum_{\ell=1}^{L_0} 
    |g_{\ell 0}|^{2}
\end{equation}
which represents the channel estimate error
relative to the average wideband received channel energy.
This error can be interpreted as the measure of predicting the channel gain at displaced locations from ray tracing at locations close to the
reference.

Fig.~\ref{fig:ecdf_lambda_rnd_28_140} plot the empirical cumulative distribution function of the error \eqref{eq:ref_err} in both \SI{28}{GHz} and \SI{140}{GHz}.
As expected, for all models, as the displacement
from the reference location is increased, the error increases
since we are trying to extrapolate the channel further from the 
reference. 
We also observe that the two-parameter estimation methods
of the reflection model, RM-RT and RM-DP,  have similar performance.
This fact shows that the two-parameter estimation methods are in agreement.

Most importantly, we see that the reflection model (either RM-DP or RM-RT)
obtains dramatically lower errors at high displacements
than the PWA or constant model.
For example, at a \SI{100}{cm} displacement,
the  median relative error of the RM is less than $10^{-2}$, thus  enabling accurate calculation of the MIMO
matrix terms with apertures of this size.  
In contrast, the relative error is $>1$ for the PWA and constant
model.  Interestingly, although the proposed
reflection model is only theoretically correct 
for fully specular reflections, we see that
low errors are obtainable even with foliage and diffraction.

\iftoggle{conference}{}{
\section{Application for Estimation LOS/NLOS MIMO Capacity}
\iftoggle{onecolumn}{  
   \begin{table}[]
    \centering
    \caption {Channel capacity estimation simulation parameters} \label{tb:parameter_table} 
    \begin{tabular}{|l|lll|}
    \hline
    Item                      & \multicolumn{3}{c|}{Value}                                           \\ \hline
    \multirow{2}{*}{Spectrum} & \multicolumn{3}{l|}{Carrier frequency: 140 GHz}                      \\ \cline{2-4} 
                              & \multicolumn{3}{l|}{Bandwidth: 2 GHz}                                \\ \hline
    Antenna Height            & \multicolumn{3}{l|}{TX \& RX: 2.49 m (central point)}                \\ \hline
    Array Size                & \multicolumn{3}{l|}{TX \& RX: 64 (8 $\times$ 8 UPA)}                 \\ \hline
    Antenna Spacing           & \multicolumn{3}{l|}{0.14 m ($\sim$65 * wavelenth)}                   \\ \hline
    Array Aperture            & \multicolumn{3}{l|}{0.98 m (Horizontal and Vertical)}                \\ \hline
    Transmit Power            & \multicolumn{3}{l|}{TX Array: 23 dBm}                                \\ \hline
    Noise Figure              & \multicolumn{3}{l|}{3 dB}                                            \\ \hline
    TX Array Orientation      & \multicolumn{3}{l|}{$[-180^\circ, 180^\circ]$ with $15^\circ$ steps} \\ \hline
    RX Array Orientation      & \multicolumn{3}{l|}{Align on boresight (face to TX)}                 \\ \hline
    Tx-Rx Distance            & \multicolumn{3}{l|}{180 meters}                                      \\ \hline
    \end{tabular}
    \end{table}
}{
   \begin{table}[]
    \centering
    \caption {Channel capacity estimation simulation parameters} \label{tb:parameter_table} 

    \begin{tabular}{@{}llll@{}}
    \toprule
    Item                      & \multicolumn{3}{c}{Value}                                           \\ \midrule
    \multirow{2}{*}{Spectrum} & \multicolumn{3}{l}{Carrier frequency: 140 GHz}                      \\
                              & \multicolumn{3}{l}{Bandwidth: 2 GHz}                                \\
    Antenna Height            & \multicolumn{3}{l}{TX \& RX: 2.49 m (central point)}                \\
    Array Size                & \multicolumn{3}{l}{TX \& RX: 64 (8 $\times$ 8 UPA)}                 \\
    Antenna Spacing           & \multicolumn{3}{l}{0.14 m ($\sim$65 * wavelength)}                   \\
    Array Aperture            & \multicolumn{3}{l}{0.98 m (Horizontal and Vertical)}                \\
    Transmit Power            & \multicolumn{3}{l}{TX Array: 23 dBm}                                \\
    Noise Figure              & \multicolumn{3}{l}{3 dB}                                            \\
    TX Array Orientation      & \multicolumn{3}{l}{$[-180^\circ, 180^\circ]$ with $15^\circ$ steps} \\
    RX Array Orientation      & \multicolumn{3}{l}{Align on boresight (face to TX)}                 \\
    Tx-Rx Distance            & \multicolumn{3}{l}{180 meters}                                      \\ \bottomrule
    \end{tabular}
    \end{table}

}

\iftoggle{onecolumn}{  
   \begin{SCfigure}
      \centering
      \includegraphics[scale=0.6]{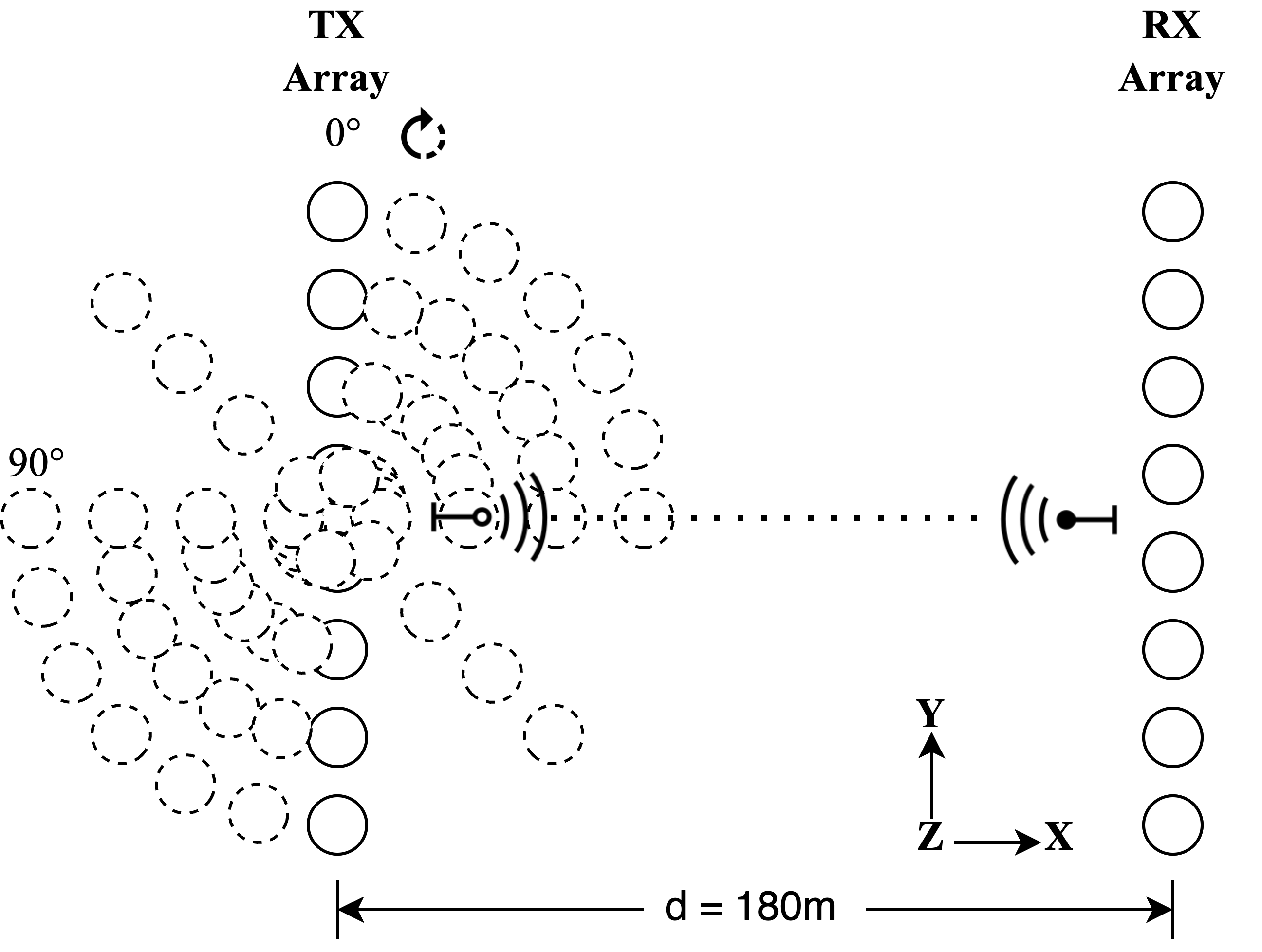}
      \caption{Illustration of TX and RX arrays orientation. The orientation of the RX array is aligned on the boresight and faces the center of the TX array. The TX array is rotated at different angles $\phi$ away from the foresight. The goal is to estimate
      the MIMO capacity as a function of the antenna orientation. We use values $\phi \in [-180^{\circ},180^{\circ}]$ with $15^{\circ}$ steps.}
      \label{fig:TXRot}
      \vspace{-4mm}
    \end{SCfigure}
}{
   \begin{figure}
      \centering
      \includegraphics[width=0.8\linewidth]{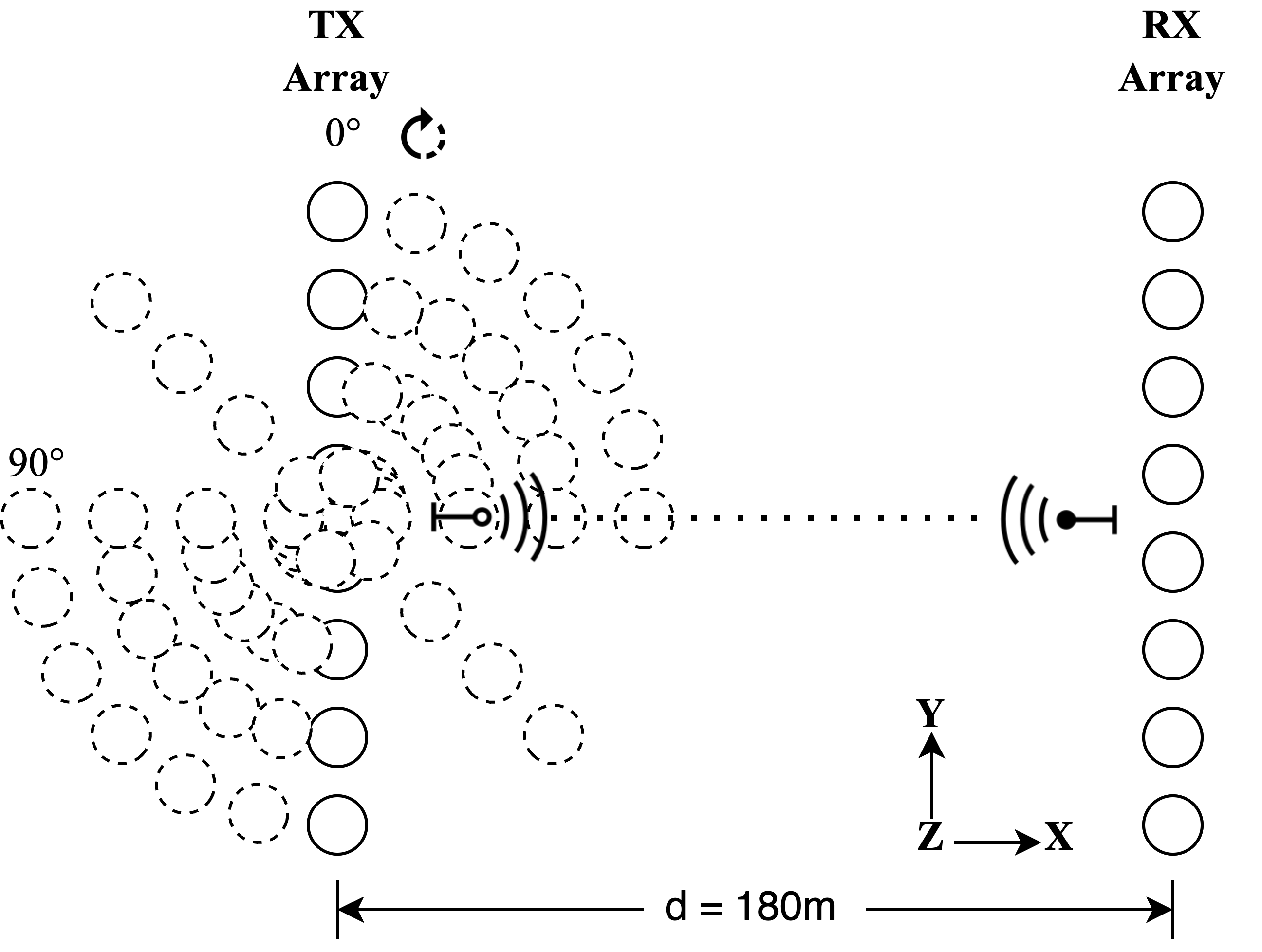}
      \caption{Illustration of TX and RX arrays orientation.
      The orientation of the RX array is aligned on the boresight and faces the center of the TX array. The TX array is rotated at different angles $\phi$ away from the boresight.  The goal is to estimate
      the MIMO capacity as a function of the antenna orientation.  We use 
      values $\phi \in [-180^{\circ},180^{\circ}]$ with $15^{\circ}$ steps.}
      \label{fig:TXRot}
    \end{figure}
}

\begin{figure*}
\footnotesize
    \centering
        \includegraphics[width=\textwidth]{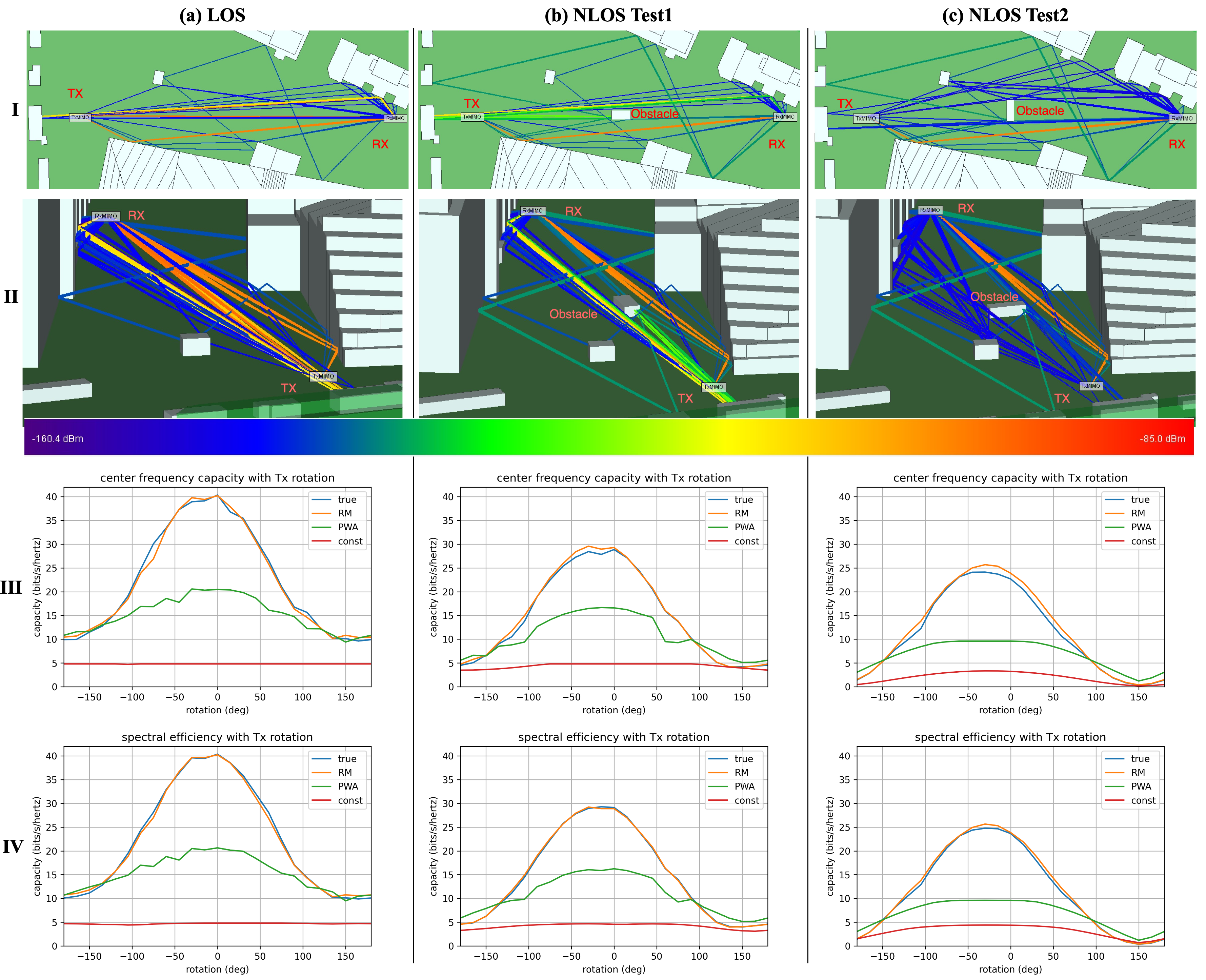}
\caption{Channel capacity simulation results and urban test scenarios at \SI{140}{GHz}. 
There are three different test scenarios shown in Column (a, b, c).
The Row I and Row II depict 2D (top-down) and 3D views of the paths between TX and RX, respectively.
In each test case, the transmitter is rotated
around 360$^\circ$.
Within each test case, the figures in Rows III and
IV show the true and estimated capacity as
a function of the TX rotation angle.
Row III shows the narrowband capacity at the center
frequency, and Row IV shows the average
spectral efficiency across the band.}
\label{fig:capacity}
\vspace{-2mm}
\end{figure*}

\iftoggle{onecolumn}{  
}{


}


\subsection{Simulation Set-Up}
We conclude with a demonstration example of how the RM model can 
be used to significantly reduce the computation time
in predicting the MIMO capacity in a wide aperture system.
The parameters of the channel capacity estimation simulation
are shown in the Table~\ref{tb:parameter_table}.

We select a single TX and RX location pair in the Beijing area
with a TX-RX separation distance of approximately $d=$\,\SI{180}{m}.
All simulations are performed at a carrier frequency of $f_0=$\,\SI{140}{GHz} and the bandwidth $B=$\,\SI{2}{GHz}
-- similar to what is being expected for sub-THz backhaul
\cite{sawaby2020fully,gougeon2020assessment,chintareddy2021preliminary}.
We then consider three conditions:
\begin{itemize}
    \item Test (a):  The  environment with no additional obstacles.
    In this case, the link from the TX-RX is LOS.
    \item Test (b):  An additional $2 \times 3 \times 4$\,\si{m^3} obstacle (similar to a car) is 
    placed to block the LOS path.  The obstacle is oriented in the $x$-axis (east-west).
     \item Test (c):  The identical set-up as Test (b), but the obstacle oriented in the $y$-axis (north south).
\end{itemize}
A top-down view of each of the test scenarios 
is shown in the middle panel of
Fig.~\ref{fig:capacity} and a 3D view is shown in the bottom panel.
Ray tracing is run between the TX-RX locations in Tests (a)--(c)
and the rays found from the ray tracing are also shown in middle and bottom panels of of
Fig.~\ref{fig:capacity}.  It can be seen that Test (a) has a LOS path while Tests (b) and (c) have only NLOS paths.
All ray tracing simulations included diffraction and foliage.
In particular, diffracted paths around the obstacles
can be seen in Tests (b) and (c) in Fig.~\ref{fig:capacity}.
Assuming an ideal planar reflector, the RM model would indeed be accurate.
However, since actual reflectors are neither infinitely large nor perfectly planar, differences between the true and RM model arise.
Nonetheless, empirical studies show that the model remains a dependable approximation even with substantial displacements.

We then place $8\times 8$ uniform planar arrays (UPAs) 
on both TX and RX sides with an array total aperture of $0.98 \times 0.98$\,\si{m}, in which case, the antenna spacing is $0.14$\,\si{m}. We adopt the gNB antenna pattern specified by 3GPP~\cite{3GPP36873}. 

The arrays are first aligned to each other so
that their bore sights are along the LOS direction 
(even in Tests (b) and (c) where the LOS path is not present).  
We then consider 
azimuth rotations $\phi$ of the TX array away from bore sight.
We use $N_{\rm ang}=24$ angular values of
$\phi \in [-180^{\circ},180^{\circ}]$ with $15^{\circ}$ steps.
The rotation is illustrated in Fig.~\ref{fig:TXRot}.

Our goal is to estimate the MIMO capacity of this link as a function
of the TX array orientation $\phi$.
This type of simulation would often occur in RF planning
since one may need to understand how to mount and orient the array
for optimal coverage.  Also, 
when serving multiple points,
the array cannot be oriented in bore sight for all RX locations.  In this case, it is valuable
to be able to predict the MIMO capacity as a function of the actual orientation.

We emphasize here that our goal here is not to make a general
statement on the capacity of wide aperture MIMO systems.
Such an analysis would require running more extensive simulations to
find the 
statistical distribution of the capacity over large numbers of TX-RX locations.
The point of this simulation is to simply illustrate how the 
RM model can be used to simplify the simulation time for one such link.

\subsection{Capacity Estimation via Exhaustive Ray Tracing}
\label{sec:cap_true}

We first consider estimating the capacity via exhaustive
ray tracing.  This method is the most exact, but also the most computationally
intensive.
For exhaustive ray tracing, we must run ray tracing between each TX and RX element in the arrays at each orientation at some reference frequency
$f_0$.  That is, between each TX element $n$ and RX element $m$,
we use ray tracing at the RF center frequency $f_0$ to find parameters 
\begin{equation} \label{eq:meas_capacity}
    (g_{\ell m n},\tau_{\ell m n}, \phi_{\ell m n}^r,\theta_{\ell mn}^r, 
        \phi_{\ell mn}^t,\theta_{\ell mn}^t), \quad
        \ell = 1,\ldots, L_{mn},
\end{equation}
where $L_{mn}$ is the number of paths between the TX element $n$
and RX element $n$ and the items in the vector in \eqref{eq:meas_capacity}
are the gain, delays, and angles of the path $\ell$ in that link.
Then, similar to the previous section, the MIMO channel matrix 
at a frequency $f$, can be estimated by
\begin{equation} \label{eq:Hmat}
    \nbH(f) = \begin{bmatrix} 
    H_{11}(f) & \dots  & H_{1N_{\rm tx}}(f)\\
    \vdots & \ddots & \vdots\\
    H_{N_{\rm rx}1}(f) & \dots  & H_{N_{\rm rx}N_{\rm tx}}(f) 
    \end{bmatrix},
\end{equation}
where
\begin{equation} \label{eq:Htrue_cap}
        H_{mn}(f) = 
        \sum_{\ell=1}^{L_{mn}} g_{\ell mn} e^{-j 2\pi (f-f_0)\tau_{\ell mn}}.
\end{equation}

The capacity can then be estimated from the MIMO channel
matrix from standard MIMO communication theory \cite{heath2018foundations},
depending on the MIMO assumptions.
For example, suppose that the transmit power is $P_{\rm tx}$,
the bandwidth is $B$, and the TX must transmit a constant
PSD, $S_{\rm tx} = P_{\rm tx}/B$.
Suppose, in addition, that the TX and RX know the MIMO
channel matrix, $\nbH(f)$, at all frequencies $f$ in the band
and perform optimal pre-coding at the TX and linear processing at the RX.  That is, the system has full CSI-T and CSI-R.
Let 
\begin{equation}
    \nbs(f) = (s_1(f),\ldots,s_r(f)),
\end{equation}
denote the singular values of $\nbH(f)$ at frequency $f$,
where $r$ is the channel rank.
Then, we can estimate the rate by

\begin{equation} \label{eq:Rtot}
    R := \int_{f_0-B/2}^{f_0+B/2} \mathrm{SE}(f)\, df
\end{equation}
where $\mathrm{SE}(f)$ is the spectral efficiency (i.e., 
rate per unit bandwidth):
\begin{equation} \label{eq:SE}
    \mathrm{SE}(f) = \max_{k=1,\ldots,r} \sum_{i=1}^k \rho \left(  
    \frac{s_i^2(f)P_{\rm tx} }{N_0 B k} \right),
\end{equation}
where $N_0$ is the noise PSD, $\rho(\gamma)$ is the spectral
efficiency per stream for an SNR $\gamma$, and 
the maximization over $k$ is to select the number of streams to use.
The formula \eqref{eq:SE} assumes that we allocate a fraction $1/k$
of the power to each stream and optimize over $k$.  This allocation
is an approximation of water-filling.
The resulting average spectral efficiency is
\begin{equation} \label{eq:SEavg}
    \overline{\mathrm{SE}} := R/B.    
\end{equation}
For the theoretical Shannon capacity, in \eqref{eq:SE},
we would use the formula:
\begin{equation}
    \rho(\gamma) = \log_2(1 + \gamma).
\end{equation}
However, to account for losses with practical codes and overhead, 
we assume a widely-used
model in 3GPP simulations \cite{mogensen2007lte}:
\begin{equation}
    \rho(\gamma) := \min\left\{ \alpha \log_2(1 + \gamma), \rho_{\rm max} \right\}
\end{equation}
where $\alpha = 0.6$ and $\rho_{\rm max}=4.8$\,\si{bps/Hz}.

The key computational challenge in the exhaustive capacity estimation is the ray tracing.
Since the arrays have $N_{\rm rx} = N_{\rm tx} = 64$
elements each, and there are $N_{\rm ang} = 24$ 
angular steps, we must run ray tracing $N$ times with
\begin{equation} \label{eq:nray_exhaustive}
    N= N_{\rm rx}N_{\rm tx}N_{\rm ang} = (64)(64)(24) \approx 98000,
\end{equation}
to extract the parameters \eqref{eq:meas_capacity} for 
all the angular steps.
While the exhaustive procedure is the most accurate, the large number of ray tracing simulations required can be computationally extremely expensive.
 
To illustrate the possible gains, Table~\ref{tb:time_comp} shows the computation time
 of each of the main steps for the exhaustive procedure and the RM-DP and RM-RT methods. 
 All times are on a machine with NVIDIA RTX 3090 and Intel i9-10900K.
 The exhaustive method requires significantly more 
 time due to the need to perform ray tracing at each
 of the $N_{\rm ang}=24$ angular rotations and for
 all the TX-RX pairs.  In contrast, the RM-DP and
 RM-RT methods require ray tracing only once or twice.
 The RM-DP and RM-DT methods require a parameter
 extraction component that is not needed for the exhaustive method -- but this step is negligible 
 in computation time.  All methods require similar 
 time to compute the channels from the rays, but
 again, this step is also small.
 Overall, RM-DP and RM-RT are 150 to 200 times
 faster than the exhaustive method.
 
Table~\ref{tb:time_comp}
 also shows the computational time 
 for the standard PWA method.  We see that
 the proposed RM-DP and RM-RT are slightly
 longer due to the computation of the distance
 with the orthogonal matrix.  However,
 the total computational time for RM-RT and RM-DP is approximately
 only 30 to 60\% more than PWA and, as we will see, offers a 
 much more accurate channel estimate.

 Of course, the absolute numbers will depend on
 the machine used.  However, given the massive
 reduction in ray tracing needed,
 the general trend will likely hold across platforms.


\begin{table}[]
    \centering
    \caption {Computational time comparison.} \label{tb:time_comp} 
    \begin{tabular}{|l|c|c|c|c|}
    \hline
    \multicolumn{1}{|c|}{}                                                & Exhaustive                                                  & PWA                                                         & RM-DP                                                        & RM-RT                                                        \\ \hline
    \begin{tabular}[c]{@{}l@{}}Ray \\ Tracing (min)\end{tabular}          & \begin{tabular}[c]{@{}c@{}}$65.4$\\ $\times24$\end{tabular} & 2.1                                                         & 4.3                                                          & 2.1                                                          \\ \hline
    \begin{tabular}[c]{@{}l@{}}Parameter \\ Estimation (sec)\end{tabular} & -                                                           & -                                                           & 0.0014                                                       & 0.0023                                                       \\ \hline
    \begin{tabular}[c]{@{}l@{}}Channel \\ Computation (sec)\end{tabular}  & \begin{tabular}[c]{@{}c@{}}$7.31$\\ $\times24$\end{tabular} & \begin{tabular}[c]{@{}c@{}}$9.59$\\ $\times24$\end{tabular} & \begin{tabular}[c]{@{}c@{}}$14.57$\\ $\times24$\end{tabular} & \begin{tabular}[c]{@{}c@{}}$15.11$\\ $\times24$\end{tabular} \\ \hline
    Total (min)                                                           & 1572.5                                                      & 5.9                                                         & 10.1                                                         & 8.1                                                          \\ \hline
    \end{tabular}
\end{table}

Note that the purpose here is not to suggest a particular
MIMO scheme.  For example, in certain scenarios,
CSI-T may not be available.  In these cases, the rate
formula may be different.  However, whatever the scheme
is used, one will similarly need to compute the channel
matrix at different array configurations, and the same
computational complexity problem will hold.
We simply select the above MIMO capacity problem
since these computational difficulties are clear to see.

\subsection{Approximate Capacity Estimation}
Similar to Section~\ref{sec:valid},
we next consider the approximate capacity 
estimation using constant, PWA, or RM channel estimates.
While these methods are approximate, the advantage is that,
instead of running $N$ ray tracing
simulations, where $N$ is given in \eqref{eq:nray_exhaustive}, we only need to 
run a \emph{single} ray tracing
simulation between a reference location 
$\nbx^t_0$ in the center of the TX array
and a reference location $\nbx^r_0$
in the center of the RX array.
The MIMO channel, with any array orientation,
can be then estimated from this ray tracing data, providing
a much more computationally efficient approach to estimating the capacity.
Our interest is in comparing the quality of this capacity estimate
for different methods.
 
The details of the process are as follows:
The ray tracing  provides
the PWA parameters \eqref{eq:param_pwa} 
between the reference TX-RX pair
$(\nbx^t_0,\nbx^r_0)$.  Similar to Section~\ref{sec:valid},
we consider three possible approximations for the MIMO channel matrices $H_{mn}(f)$: A constant, PWA, and RM model.  Each model provides an estimate
for the channel, $\widehat{H}_{mn}(f)$.
The formula for these estimates is similar to 
Section~\ref{sec:valid}.  For example, the PWA 
and RM model provide
estimates $\widehat{d}_{\ell mn}$ of the delay
from TX element $n$ to RX element $m$ on path $\ell$.
The delay estimates can then be used in a formula similar to \eqref{eq:Hhat} to estimate $\widehat{H}_{mn}(f)$.
The channel estimates $\widehat{H}_{mn}$ can then be used in place 
of the true coefficients $H_{mn}(f)$ in \eqref{eq:Htrue_cap}.  Then, the achievable rate $R$
in \eqref{eq:Rtot} can be estimated using the channel estimates to obtain an approximation of $R$.

\subsection{Results}
The third and fourth rows of Fig.~\ref{fig:capacity}
show the  MIMO channel center frequency capacity and spectral efficiency, $R/B$, as a function of the angle $\phi$,
where the true channel,
computed from exhaustive ray tracing, is used. 
And we simplify the integral in \eqref{eq:Rtot} for computing $R$ by summation over ten uniformly distributed frequencies within the bandwidth.
As expected, the true channel capacity of the LOS link is greater than that of NLOS links.  Also, the LOS capacity is maximized
by when the arrays are pointed at bore sight.
For the NLOS cases, the optimal pointing
angle is slightly off bore sight to capture
dominant reflections.

Also plotted in the fourth row of Fig.~\ref{fig:capacity} is the capacity
estimate using different channel estimates for different methods.
We see that the capacity estimate by the
RM model is close to the true capacity.
Indeed, the overall error of the RM's estimation of channel capacity is less than $5\%$.
In contrast, the PWA and constant model grossly under-predict the capacity.

\iftoggle{onecolumn}{  
}{
}


Overall, we see that the RM model can provide an estimate
of the capacity of a wide aperture array \SI{140}{GHz} 
system in a complex urban environment with reflections.  Specifically, the RM model matches
the capacity estimated via exhaustive ray tracing,
but comes with a dramatically lower simulation time.
While exhaustive ray tracing requires 
one ray tracing simulation between every TX-RX element
and every array configuration, RM requires a single
ray tracing simulation.  The PWA and constant
models also save the ray tracing, but are grossly
inaccurate for wide aperture arrays.

    


}

\section{Conclusions}
Near-field communication
is a promising technology for systems in the 
mmWave and THz bands.  However, accurate assessment
of near-field communications requires
channel models that can capture the spherical
nature of the wavefront of each path, a feature
not accounted for in most models today that
use planar approximations of waves. This paper has presented a simple parametrization for 
multi-path wireless channels that correctly describes
the spherical nature of each wavefront.
Interestingly, the parametrization
requires only two additional parameters relative
to standard plane wave models.  Moreover,
we have provided a computationally simple
algorithm to extract the parameters
from ray tracing.

The model is based on image theory and is 
fully accurate under the assumption
of planar, infinite surface reflections.
Moreover, our simulations show that the proposed reflection model delivers a high accuracy
over wide apertures, even when these exact conditions are not met.  In particular, the models are significantly more accurate than models
based on plane wave approximations.  The technique is precise while substantially decreasing
the simulation duration in contrast to the comprehensive ray tracing approach.

Going forward, the method can greatly enhance
the evaluation of near-field communications in site-specific settings. In this paper, we have
demonstrated the method for evaluation of 
mmWave  and sub-THz wide-aperture MIMO backhaul links in a site-specific setting. 

A natural 
next step is to develop statistical channel 
models, such as those used by 3GPP \cite{3GPP38901},
or machine learning methods \cite{XiaRanMez2020, HuAsilomar2022Multi},
that describe the distribution of these parameters
in common environments.

\iftoggle{conference}{}
{
\appendices 
\section{Proof of 
Theorem~\ref{thm:path_dist_orthogonal}}
\label{sec:proof_theorem1}


Write the path's route 
as a sequence of $K-1$ interactions as in \eqref{eq:path_sequence}.  Let $S_k$ denote the $k$-th
reflecting plane.  
The initial transmitter point $\nbx^t$ can be reflected
across the surface $S_1$ to obtain an image that we will denote
$\nbz^1$.  This image point can in turn be reflected to
create a second image $\nbz^2$.  After $K-1$ reflections,
we obtain a final image point $\nbz^{K-1}$.

The method of images states that the total distance of the reflected path is equal to 
the LOS distance from the final reflected 
image point $\nbz^{K-1}$ to the receiver $\nbx^r$.
Hence,
\begin{equation} \label{eq:dproof1}
    d(\nbx^r,\nbx^t) = \| \nbx^r - \nbz^{K-1} \|.
\end{equation}
Therefore, \eqref{eq:dnlosUg} will be proven if we can show
\begin{equation} \label{eq:zUg}
    \nbz^{K-1} = \nbU \nbx^t + \nbg,
\end{equation}
for some orthogonal matrix $\nbU$ and vector $\nbg$.
That is, the image point is a rotation and translation of the 
original transmitted point.

Finding the matrix $\nbU$ and vector $\nbg$ in \eqref{eq:zUg} 
is a matter of simple geometry.  
We will walk through the details since this process will also 
show how to numerically compute the parameters from the 
route sequence \eqref{eq:path_sequence}.

First, since each surface is a plane, the surface can be represented as:
\begin{equation} \label{eq:Sk}
    S_k = \left\{ \nbx ~|~ (\nbu^k)^\intercal \nbx = b^k
    \right\},
\end{equation}
for some unit vector $\nbu^k$ and constant $b^k$.
To compute the normal vector,  let $\nbv^k$ be the unit vector
of the $k$-th step in the route:
\begin{equation} \label{eq:dir_vec}
    \nbv^k = \frac{\nbx^{k} - \nbx^{k-1}}{\|\nbx^{k} - \nbx^{k-1}\|}, \quad k=1,\ldots,K.
\end{equation} 
Then, the normal vector for $S_k$ is given by:
\begin{equation} \label{eq:norm_vec}
    \nbu^k := \frac{\nbv^{k+1} - \nbv^k}{\|\nbv^{k+1} - \nbv^k\|}, \quad k=1,\ldots,K-1.
\end{equation}
Also, since we know $\nbx^k$ is in the plane $S_k$ in \eqref{eq:Sk},
the intercept must be given by:
\begin{equation} \label{eq:intercept}
    b^k =(\nbu^k)^\intercal \nbx^k.
\end{equation}
Since the image point $\nbz^k$ is the reflection of $\nbz^{k-1}$ around $S_k$,
the two points are related by:
\begin{align} 
    \nbz^k &= \nbx^{k-1} - 2\nbu^k ((\nbu^k)^\intercal \nbz^{k-1} - b^k) \nonumber \\
        &= \nbV_k\nbz^{k-1} + \nbc_k \label{eq:zrec}
\end{align}
where
\begin{equation} \label{eq:Vkck}
    \nbV_k = \nbI - 2\nbu^k(\nbu^k)^\intercal,
    \quad
    \nbc_k = 2b^k\nbu^k.
\end{equation}
The recursion \eqref{eq:zrec} should be initialized with  $\nbz^0 = \nbx^t$.
Solving \eqref{eq:zrec}, we obtain
\begin{equation}
    \nbz^k = \nbU_k \nbx^t + \nbg^k,
\end{equation}
where  
\begin{equation} \label{eq:Uprod}
    \nbU_k = \prod_{i=1}^k \nbV_i,
\end{equation}
and $\nbg^k$ satisfies the recursions
\begin{equation} \label{eq:grec}
    \nbg^k = \nbc^k + \nbV_k\nbg^{k-1}, \quad k=1,\ldots,K-1,
\end{equation}
with the initial condition $\nbg^0=0$.
Iterating through \eqref{eq:zrec}, we obtain that
the final image point is given by \eqref{eq:zUg} with
\begin{equation}
    \nbU = \prod_{k=1}^{K-1} \nbV_k, \quad \nbg = \nbg^{K-1}.
\end{equation}
Also, each matrix $\nbV_k$ in \eqref{eq:Vkck} is orthogonal.  In fact, it is a Housholder
matrix.  Since $\nbU$ in \eqref{eq:Uprod} is the product of these matrices, 
$\nbU$ is also orthogonal.  This completes the proof.

\section{Proof of Theorem~\ref{thm:path_dist_param}}
\label{sec:proof_path_dist_param}
From Theorem~\ref{thm:path_dist_orthogonal}, we know the
distance function $d(\nbx^r,\nbx^t)$ can be written 
as \eqref{eq:dnlosUg} for some matrix
$\nbU$ and translation vector $\nbg$.
So, we can prove the theorem if we can rewrite 
\eqref{eq:dnlosUg} as \eqref{eq:dnlos}.
Let $\nbz_0^t$ denote 
the reflected image of the TX reference $\nbx^t_0$:
\begin{equation}
    \nbz_0^t := \nbU\nbx^t_0 + \nbg,
\end{equation}
and let $\nbd_0$ denote 
the vector from the RX to the reflection of the
TX:
\begin{equation} \label{eq:dreflect}
    \nbd_0 := \nbx^r_0 - \nbz^t_0 = \nbx^r_0 - \nbU \nbx^t_0 -\nbg.
\end{equation}
Then, for any points $\nbx^t$ and $\nbx^r$, we can subtract off $\nbx^t_0$ and $\nbx^r_0$ to rewrite  \eqref{eq:dnlosUg} as
\begin{align} 
    \MoveEqLeft d(\nbx^r,\nbx^t) = \Bigl\| 
         \nbx^r-\nbx^r_0 - \nbU (\nbx^t - \nbx^t_0) + \nbd_0 \Bigr\|, \label{eq:dproof2}
\end{align}
Next,  let
\begin{equation} \label{eq:taupf}
    \tau = \frac{1}{c}\|\nbd_0\|,
\end{equation}
which represents the time of flight from the reference RX
to the reflected image of the  TX.
Since $\nbd_0 \in \R^3$ with $\|\nbd_0\| = c\tau$,
we can write $\nbd_0$ in spherical coordinates:
\begin{equation} \label{eq:dspherical}
    \nbd_0 = c\tau (\cos \theta^r \cos \phi^r, \cos \theta^r \sin \phi^r, -\sin \theta^r).
\end{equation}
for angles $\phi^r$ and $\theta^r$.  The spherical coordinates \eqref{eq:dspherical}
can also be written as:
\begin{equation} \label{eq:angrpf}
    \nbR_y(\theta^r)\nbR_z(-\phi^r)\nbd_0 = c\tau\nbe_x,
\end{equation}
where $\nbe_x=(1,0,0)$ is the unit vector in the $x$-direction.
Substituting \eqref{eq:angrpf} into
\eqref{eq:dproof2}, we obtain
\iftoggle{onecolumn}{  
   \begin{align} 
        \MoveEqLeft d(\nbx^r,\nbx^t) \stackrel{(a)}{=} \Bigl\| 
             \nbR_y(\theta^r)\nbR_z(-\phi^r)
              \times \left[ (\nbx^r
             -\nbx^r_0) 
             - \nbU (\nbx^t - \nbx^t_0) + \nbd_0 \right] \Bigr\| \nonumber \\
        & \qquad \stackrel{(b)}{=}
         \Bigl\| 
             \nbR_y(\theta^r)\nbR_z(-\phi^r)(\nbx^r
             -\nbx^r_0)  + \nbW (\nbx^t - \nbx^t_0) + c\tau\nbe_x \Bigr\|,
        \label{eq:dproof3}
    \end{align}
}{
   \begin{align} 
        \MoveEqLeft d(\nbx^r,\nbx^t) \stackrel{(a)}{=} \Bigl\| 
             \nbR_y(\theta^r)\nbR_z(-\phi^r)
             \nonumber \\
        & \qquad \times \left[ (\nbx^r
             -\nbx^r_0) 
             - \nbU (\nbx^t - \nbx^t_0) + \nbd_0 \right] \Bigr\| \nonumber \\
        & \stackrel{(b)}{=}
         \Bigl\| 
             \nbR_y(\theta^r)\nbR_z(-\phi^r)(\nbx^r
             -\nbx^r_0) \nonumber \\
        & \qquad + \nbW (\nbx^t - \nbx^t_0) + c\tau\nbe_x \Bigr\|,
        \label{eq:dproof3}
    \end{align}
}
where the first step (a) follows from 
\eqref{eq:dproof2} and the fact that
$\nbR_y(\theta^r)\nbR_z(-\phi^r)$
is a rotation matrix that does not 
change distance,
and, in step (b), we define
\begin{equation} \label{eq:Wdef}
    \nbW := 
    - \nbR_y(\theta^r)\nbR_z(-\phi^r)\nbU.
\end{equation}
Also, each Householder matrix $\nbV_k$ in \eqref{eq:Vkck} is orthogonal with 
determinant,  $\mathrm{det}(\nbV_k)=-1$.
Hence, the determinant $\nbU$ in \eqref{eq:Uprod} is:
\begin{equation} \label{eq:detU}
    \mathrm{det}(\nbU) = (-1)^{K-1}.
\end{equation}
Taking the determinant of the product \eqref{eq:Wdef},
\iftoggle{onecolumn}{  
   \begin{align}
        \mathrm{det}(\nbW) &= 
        \mathrm{det}\left[ 
         - \nbR_y(\theta^r)\nbR_z(-\phi^r)\nbU
         \right] 
         = (-1)^2 
         \mathrm{det}\left[ 
         \nbR_y(\theta^r)\nbR_z(-\phi^r)
         \right]\mathrm{det}(\nbU) 
         = (-1)^{K+3}
         = s,
    \end{align}
}{
   \begin{align}
        \mathrm{det}(\nbW) &= 
        \mathrm{det}\left[ 
         - \nbR_y(\theta^r)\nbR_z(-\phi^r)\nbU
         \right] \nonumber \\
         &= (-1)^2 
         \mathrm{det}\left[ 
         \nbR_y(\theta^r)\nbR_z(-\phi^r)
         \right]\mathrm{det}(\nbU) \nonumber \\
         & = (-1)^{K+3}
         = s,
    \end{align}
}
where $s=\pm 1$ as defined in \eqref{eq:sdef},
and we have used the fact that the determinant
of rotation matrices is one.
Multiplying by the matrix $\nbQ_z(s)$ defined
in \eqref{eq:Qzdef},
we obtain
\begin{align}
    \mathrm{det}\left[ \nbQ_z(s) \nbW \right]
    = 
    \mathrm{det}(\nbQ_z(s)) \mathrm{det}(\nbW)= s^2 = 1. 
\end{align}
Hence $\nbQ_z(s)\nbW$ is an orthogonal 
matrix with determinant of one
in $\R^{3 \times 3}$.  That is, 
the matrix is in the special orthogonal 
group of rotations $SO(3)$.  
Any such matrix can be parameterized by three rotations:
\begin{equation} \label{eq:VprodR}
    \nbQ_z(s) \nbW 
    = \nbR_x(\gamma^t)\nbR_y(\theta^t)\nbR_z(-\phi^t).
\end{equation}
Since $\nbQ_z(s)^2 = \nbI$,
\begin{equation} \label{eq:VprodR1}
    \nbW
    = \nbQ_z(s)\nbR_x(\gamma^t)\nbR_y(\theta^t)\nbR_z(-\phi^t).
\end{equation}
Substituting \eqref{eq:VprodR1} into
\eqref{eq:dproof3} proves
\eqref{eq:dnlos}.

It remains to show that parameters $(\theta^r,\phi^r,\theta^t,\phi^t)$ match
those in the PWA model.  This equivalency
is shown in Appendix~\ref{sec:proof_equiv}.

\section{Equivalency of the RM and PWA Parameters} \label{sec:proof_equiv}
Let
\begin{equation} \label{eq:ang_rm}
    (\theta^r,\phi^r,\theta^t,\phi^t)   
\end{equation}
be the angles for the RM model
derived in the Appendix~\ref{sec:proof_path_dist_param}.
We need to show that these angles
are identical to the angles in the 
PWA model.  To this end, let
$\nbu^r$ and $\nbu^t$ be the
direction vectors \eqref{eq:usph},
computed from these RM angles \eqref{eq:ang_rm}:
\begin{subequations} \label{eq:usph_equiv}
\begin{align}
    \nbu^r &= (\cos(\phi^r)\cos(\theta^r), 
    \sin(\phi^r)\cos(\theta^r), \sin(\theta^r)) \\
    \nbu^t &= (\cos(\phi^t)\cos(\theta^t), 
    \sin(\phi^t)\cos(\theta^t), \sin(\theta^t)),
\end{align}
\end{subequations}
where, similar to Appendix~\ref{sec:proof_path_dist_param},
we have dropped the dependence on $\ell$
to simplify the notation.
We will show that the direction vectors
\eqref{eq:usph_equiv} satisfy the directional
derivative property \eqref{eq:uderiv}:
\[
(\nbu^r)^\intercal = -\frac{\partial d_\ell(\nbx^r_0,\nbr^t_0)}{\partial \nbx^r},
    \quad
    (\nbu^r)^\intercal = -\frac{\partial d_\ell(\nbx^r_0,\nbr^t_0)}{\partial \nbx^t}.
\]
     
\begin{equation} \label{eq:uderiv_equiv}
     (\nbu^r)^\intercal = -\frac{\partial d_\ell(\nbx^r_0,\nbx^t_0)}{\partial \nbx^r},
    \quad
    (\nbu^r)^\intercal = -\frac{\partial d_\ell(\nbx^r_0,\nbx^t_0)}{\partial \nbx^t}.
\end{equation}

Hence the directions \eqref{eq:usph_equiv}
must match the PWA directions, and therefore,
so must the angles.

To prove \eqref{eq:uderiv_equiv},
write the distance in \eqref{eq:dnlos}
as:
\begin{equation} \label{eq:dfunc}
    d(\nbx^r,\nbx^t) = f(\nbz),   
\end{equation}
where
\begin{equation} \label{eq:fdef}
    f(\nbz) := \| c\tau \nbe_x + \nbz \|,
\end{equation}
and
\iftoggle{onecolumn}{  
   \begin{align} \label{eq:dzproof}
        \MoveEqLeft \nbz  = \nbR_y(\theta^r)\nbR_z(-\phi^r)(\nbx^r_0-\nbx^r)
             + \nbQ_z(s)\nbR_x(\gamma^t)
           \nbR_y(\theta^t)\nbR_z(-\phi^t)(\nbx^t_0-\nbx^t).
    \end{align} 
}{
   \begin{align} \label{eq:dzproof}
        \MoveEqLeft \nbz  = \nbR_y(\theta^r)\nbR_z(-\phi^r)(\nbx^r_0-\nbx^r)
            \nonumber \\
           & + \nbQ_z(s)\nbR_x(\gamma^t)
           \nbR_y(\theta^t)\nbR_z(-\phi^t)(\nbx^t_0-\nbx^t).
    \end{align}
}
Then,
\iftoggle{onecolumn}{  
   \begin{align}
        \MoveEqLeft \frac{\partial d_\ell(\nbx^r_0,\nbx^t_0)}{\partial \nbx^r}
        \stackrel{(a)}{=} \left.\frac{\partial f(\nbz)}{\partial \nbz}\right|_{\nbz = \boldsymbol{0}}
        \left.\frac{\partial \nbz}{\partial \nbx^r}\right|_{\nbx^r = \nbx^r_0} 
        \stackrel{(b)}{=}  \nbe_x^\intercal 
        \left.\frac{\partial \nbz}{\partial \nbx^r}\right|_{\nbx^r = \nbx^r_0} 
        \stackrel{(c)}{=}  -\nbe_x^\intercal 
        \nbR_y(\theta^r)\nbR_z(-\phi^r) 
        \stackrel{(d)}{=} -(\nbu^r)^\intercal,
        \label{eq:urderiv}
    \end{align}
}{
   \begin{align}
        \MoveEqLeft \frac{\partial d_\ell(\nbx^r_0,\nbx^t_0)}{\partial \nbx^r}
        \stackrel{(a)}{=} \left.\frac{\partial f(\nbz)}{\partial \nbz}\right|_{\nbz = \boldsymbol{0}}
        \left.\frac{\partial \nbz}{\partial \nbx^r}\right|_{\nbx^r = \nbx^r_0} 
        \nonumber \\
        &\stackrel{(b)}{=}  \nbe_x^\intercal 
        \left.\frac{\partial \nbz}{\partial \nbx^r}\right|_{\nbx^r = \nbx^r_0} 
        \nonumber \\
         &\stackrel{(c)}{=}  -\nbe_x^\intercal 
        \nbR_y(\theta^r)\nbR_z(-\phi^r) 
        \stackrel{(d)}{=} -(\nbu^r)^\intercal,
        \label{eq:urderiv}
    \end{align}
}
where (a) follows from \eqref{eq:dfunc}
and chain rule;
(b) follows from taking the derivative
of $f(\nbz)$ in \eqref{eq:fdef};
(c) follows from taking the derivative
of $\nbz$ in \eqref{eq:dzproof};
and (d) follows from applying the formulae
for the rotation matrices in 
\eqref{eq:rotmatrix} and 
the definition of $\nbu^r$ in
\eqref{eq:usph_equiv}.  Thus,
\eqref{eq:urderiv} proves the first equation
in \eqref{eq:uderiv_equiv}.
The derivative with respect to $\nbx^t$
is proven similarly.

\section{Estimation via Displaced Pairs} \label{sec:displaced_details}

The details of the RM-DP fitting procedure are as follows:  We assume we have PWA parameters \eqref{eq:meas_rmdp}
between $M+1$ TX-RX pairs,
$(\nbx^t_m,\nbx^r_m)$, $m=0,\ldots,M$.
As mentioned in Section~\ref{sec:rmdp},
the pair $(\nbx^t_0,\nbx^r_0)$ is the \emph{reference
pair} and $(\nbx^t_0,\nbx^r_0)$, $m=1,\ldots,M$,
are the $M$ \emph{displaced  pairs}.
The goal is to determine the RM parameters
\eqref{eq:param_high} at the reference pair
$(\nbx^t_0,\nbx^r_0)$.
We can obtain most of the RM model parameters from 
the PWA parameters at the reference pair $(\nbx^t_0,\nbx^r_0)$.
Specifically, we set the number of paths at the reference
pair to $L=L_0$, and for each path $\ell$, we set:
\begin{subequations} \label{eq:pwa_rm}
\begin{align}
    g_\ell &= g_{\ell 0}, \quad
    \tau_\ell = \tau_{\ell 0}, \\
    \phi_{\ell}^r &= \phi_{\ell 0}^r, \quad
    \theta_{\ell}^r = \theta_{\ell 0}^r, \\
    \phi_{\ell}^t &= \phi_{\ell 0}^t, \quad
    \theta_{\ell}^t = \theta_{\ell 0}^t, 
\end{align}
\end{subequations}
The only parameters in the RM model
that need to be determined are the binary variable
$s_\ell$ and angle $\gamma_\ell$.
each path $\ell$.



We proceed in two phases:  Path matching and angle solving.

\noindent 
\paragraph*{Path matching}
If the displaced locations are close to the reference location,
the number of paths should be the same, and the paths should approximately 
agree except
for the change in the path distance.  However, the ray tracing generally
outputs paths in an arbitrary order.  So, we first perform a 
heuristic \emph{path matching} as follows.   Let 
\iftoggle{onecolumn}{  
   \begin{align}
        \MoveEqLeft D_m(\ell,\ell') :=  c_0 \left[ |\phi_{\ell 0}^r - \phi_{\ell' m}^r|
        + |\phi_{\ell 0}^t - \phi_{\ell' m}^t| \right] 
        + 
        c_1 \left[ |\theta_{\ell 0}^r - \theta_{\ell' m}^r|
        + |\theta_{\ell 0}^t - \theta_{\ell' m}^t| \right], 
    \end{align}
}{
   \begin{align}
        \MoveEqLeft D_m(\ell,\ell') :=  c_0 \left[ |\phi_{\ell 0}^r - \phi_{\ell' m}^r|
        + |\phi_{\ell 0}^t - \phi_{\ell' m}^t| \right] \nonumber \\
        &+ 
        c_1 \left[ |\theta_{\ell 0}^r - \theta_{\ell' m}^r|
        + |\theta_{\ell 0}^t - \theta_{\ell' m}^t| \right], 
    \end{align}
}
which represents a distance between the parameters for the path $\ell$
in the reference pair and the path $\ell'$ in the displaced pair $m$.
The coefficients $c_i$ are weighting parameters that we take as
\[
    c_0 = c_1 = \frac{1}{180^\circ}, \quad
\]
To match the paths, we then perform the following recursion:
\begin{subequations} \label{eq:matchrec}
\begin{align}
    \sigma_m(\ell) &= \argmin_{\ell' \not \in I_\ell} D_m(\ell,\ell'), \\
     I_{m,\ell+1} &= I_{m,\ell} \cup \sigma_m(\ell).
\end{align}
\end{subequations}
which is initialized with $I_{m,0} = \varnothing$.  
For each path $\ell$ between the reference 
TX-RX pair $(\nbx^t_0,\nbx^r_0)$, 
the recursion 
\eqref{eq:matchrec} finds a closest path
$\ell' = \sigma_m(\ell)$ in the displaced pair $(\nbx^t_m,\nbx^r_m)$.
The recursion is performed from the strongest path
to the weakest path, meaning they are sorted in descending
order of $|g_{\ell 0}|$.  This sorting ensures that the
the strongest paths are given the highest priority
in the matching.

After the path matching, we reorder the paths in all
the displaced pairs so that in the new order 
path $\ell$ in reference pair $m$
corresponds to the previous path index $\sigma_m(\ell)$.

\noindent
\paragraph*{Angle Solving}



After the path matching is performed we can solve for the binary variable $s_\ell$ and angle $\gamma_\ell$
for each path as follows.
Using the RM parameters at the reference location
$m=0$, the distance along path $\ell$ between
any TX-RX pair $(\nbx^t,\nbx^r)$  is:
\begin{align} 
    \MoveEqLeft
        d_\ell(\nbx^r,\nbx^t) = \Bigl\| 
        c\tau_{\ell 0}\nbe_x +  \nbR_y(\theta^r_{\ell 0})\nbR_z(-\phi^r_{\ell 0})(\nbx^r_0-\nbx^r)
        \nonumber \\
       & + \nbQ_z(s_{\ell 0})\nbR_x(\gamma^t_{\ell 0})
       \nbR_y(\theta^t_{\ell 0})\nbR_z(-\phi^t_{\ell 0})(\nbx^t_0-\nbx^t) 
    \Bigr\|.
    \label{eq:dnlos_fit1}
\end{align}
We also know that the propagation delay, $\tau_{\ell m}$,
along the path $\ell$ between $\nbx^t_m$ to $\nbx^r_m$ is:
\begin{equation} \label{eq:taud_fit}
    c\tau_{\ell m} = d_\ell(\nbx^r_m, \nbx^t_m).
\end{equation}
Combining \eqref{eq:dnlos_fit1} and \eqref{eq:taud_fit}, 
we have
\iftoggle{onecolumn}{  
   \begin{align} 
        \MoveEqLeft 
        (c\tau_{\ell m})^2 = d^2_\ell(\nbx^r_m,\nbx^t_m) 
        = \left\| c\tau_{\ell 0} \nbe_x + \nba^r_{\ell m} 
        + \nbQ_z(s_{\ell 0})\nbR_x(\gamma^t_{\ell 0})\nba^t_{\ell m} \right\|^2,
        \label{eq:dsqfit1}
    \end{align}
}{
   \begin{align} 
        \MoveEqLeft 
        (c\tau_{\ell m})^2 = d^2_\ell(\nbx^r_m,\nbx^t_m) 
        \nonumber \\
        &= \left\| c\tau_{\ell 0} \nbe_x + \nba^r_{\ell m} 
        + \nbQ_z(s_{\ell 0})\nbR_x(\gamma^t_{\ell 0})\nba^t_{\ell m} \right\|^2,
        \label{eq:dsqfit1}
    \end{align}
}
where
\begin{subequations} \label{eq:adeffit}
\begin{align}
    \nba^r_{\ell m} &:= \nbR_y(\theta^r_{\ell 0})\nbR_z(-\phi^r_{\ell 0})
    (\nbx_0^r - \nbx_m^r), \\
    \nba^t_{\ell m} &:= \nbR_y(\theta^t_{\ell 0})\nbR_z(-\phi^t_{\ell 0})
    (\nbx_0^t - \nbx_m^t).
\end{align}
\end{subequations}
Expanding the square in \eqref{eq:dsqfit1} we obtain:
\iftoggle{onecolumn}{  
   \begin{align}  
    \MoveEqLeft 
        (c\tau_{\ell m})^2
        = (c\tau_{\ell 0})^2 + \|\nba^r_{\ell m}\|^2 + \|\nba^t_{\ell m}\|^2
        +2(c\tau_{\ell 0})\nbe_x^\intercal\nba^r_{\ell m} \nonumber \\
        & + 2(c\tau_{\ell 0})\nbe_x^\intercal
        \nbQ_z(s_{\ell 0})\nbR_x(\gamma^t_{\ell 0})\nba^t_{\ell m} 
        + 2(\nba^r_{\ell m})^\intercal
        \nbQ_z(s_{\ell 0})\nbR_x(\gamma^t_{\ell 0})\nba^t_{\ell m}. \label{eq:dsqfit2}
    \end{align} 
}{
   \begin{align}  
    \MoveEqLeft 
        (c\tau_{\ell m})^2
        = (c\tau_{\ell 0})^2 + \|\nba^r_{\ell m}\|^2 + \|\nba^t_{\ell m}\|^2
        +2(c\tau_{\ell 0})\nbe_x^\intercal\nba^r_{\ell m} \nonumber \\
        & + 2(c\tau_{\ell 0})\nbe_x^\intercal
        \nbQ_z(s_{\ell 0})\nbR_x(\gamma^t_{\ell 0})\nba^t_{\ell m} \nonumber \\    
        & + 2(\nba^r_{\ell m})^\intercal
        \nbQ_z(s_{\ell 0})\nbR_x(\gamma^t_{\ell 0})\nba^t_{\ell m}. \label{eq:dsqfit2}
    \end{align}
}
From \eqref{eq:adeffit}, we have
\begin{subequations} \label{eq:anorm}
\begin{align}
    \|\nba^r_{\ell m}\|^2 &= \|\nbx_0^r - \nbx_m^r \|^2 \\
    \|\nba^t_{\ell m}\|^2 &= \|\nbx_0^t - \nbx_m^t\|^2,
\end{align}
\end{subequations}
since the rotation matrices do not change the norm.
Also, combining \eqref{eq:adeffit}  and \eqref{eq:usph} we have that
\begin{subequations} \label{eq:aprod}
\begin{align}
    \nbe_x^\intercal \nba^r_{\ell m} &= (\nbu^r_{\ell})^\intercal (\nbx_0^r - \nbx_m^r) \\
    \nbe_x^\intercal \nba^t_{\ell m} &= (\nbu^t_{\ell})^\intercal (\nbx_0^t - \nbx_m^t),
\end{align}
\end{subequations}
where $\nbu^r_{\ell}$ and $\nbu^t_{\ell}$
are unit vectors in the directions of arrival and departure
at the reference locations at path $\ell$:
\begin{subequations} \label{eq:usph_fit}
\begin{align}
    \nbu^r_{\ell} &= (\cos(\phi^r_{\ell 0})\cos(\theta^r_{\ell 0}), 
    \sin(\phi^r_{\ell 0})\cos(\theta^r_{\ell 0}), \sin(\theta^r_{\ell 0})) \\
    \nbu^t_{\ell} &= (\cos(\phi^t_{\ell 0})\cos(\theta^t_{\ell 0}), 
    \sin(\phi^t_{\ell 0})\cos(\theta^t_{\ell 0}), \sin(\theta^t_{\ell 0})),
\end{align}
\end{subequations}
Substituting \eqref{eq:anorm} and \eqref{eq:aprod} into 
\eqref{eq:dsqfit2}, we obtain:
\begin{equation} 
    d^2(\nbx^r_m,\nbx^t_m) = G_{\ell m} 
    + 2(\nba^r_{\ell m})^\intercal
    \nbQ_z(s_{\ell 0})\nbR_x(\gamma^t_{\ell 0})\nba^t_{\ell m},
\end{equation}
where
\iftoggle{onecolumn}{  
   \begin{align} \label{eq:dsqfit3}
        G_{\ell m} &:= (c\tau_{\ell 0})^2 + \|\nbx_0^r - \nbx_m^r \|^2+ \|\nbx_0^t - \nbx_m^t \|^2 
        + 2c\tau_{\ell 0} \left[ (\nbu^r_\ell)^\intercal (\nbx_0^r - \nbx_m^r) + 
        (\nbu^t_\ell)^\intercal (\nbx_0^t - \nbx_m^t) \right] \nonumber \\
        &= -(c\tau_{\ell 0})^2 + \|\nbx_0^r - \nbx_m^r + 
        c\tau_{\ell 0} \nbu^r_{\ell}\|^2
         + \|\nbx_0^t - \nbx_m^t + c\tau_{\ell 0} \nbu^t_{\ell} \|^2.
    \end{align}
}{
   \begin{align} \label{eq:dsqfit3}
        G_{\ell m} &:= (c\tau_{\ell 0})^2 + \|\nbx_0^r - \nbx_m^r \|^2+ \|\nbx_0^t - \nbx_m^t \|^2 
        \nonumber \\
        & + 2c\tau_{\ell 0} \left[ (\nbu^r_\ell)^\intercal (\nbx_0^r - \nbx_m^r) + 
        (\nbu^t_\ell)^\intercal (\nbx_0^t - \nbx_m^t) \right] \nonumber \\
        &= -(c\tau_{\ell 0})^2 + \|\nbx_0^r - \nbx_m^r + 
        c\tau_{\ell 0} \nbu^r_{\ell}\|^2 \nonumber \\
        & \quad + \|\nbx_0^t - \nbx_m^t + c\tau_{\ell 0} \nbu^t_{\ell} \|^2.
    \end{align}
}
Also, write
\begin{equation}
    \nba^r_{\ell m} = (a_{1m}^r,a_{2m}^r,a_{3m}^r),
    \quad
    \nba^t_{\ell m} = (a_{1m}^t,a_{2m}^t,a_{3m}^t),
\end{equation}
where we drop the dependence on the $\ell$ to simplify
the notation.
Then from \eqref{eq:adeffit} and 
\eqref{eq:rotmatrix}, we have:
\begin{align}
    \MoveEqLeft 
    2(\nba^r_{\ell m})^\intercal
    \nbQ_z(s_{\ell 0})\nbR_x(\gamma^t_{\ell 0})\nba^t_{\ell m} = 
        2a_{1m}^ra_{1m}^t \nonumber \\
        & + 2(a_{2m}^ra_{2m}^t + s_{\ell 0}a_{3m}^ra_{3m}^t)\cos \gamma^t_{\ell 0} \nonumber \\
        & + 2(s_{\ell 0}a_{3m}^ra_{2m}^t - a_{2m}^ra_{3m}^t)\sin \gamma^t_{\ell 0}.
\end{align}
We can rewrite \eqref{eq:dsqfit2} as 
\begin{equation} \label{eq:linabc}
    C_{\ell m} = A_{\ell m}(s_{\ell 0}) x_\ell + B_{\ell m}(s_{\ell 0}) y_\ell,
\end{equation}
where 
\[
    A_m(s) := 2(a_{2m}^ra_{2m}^t + sa_{3m}^ra_{3m}^t)
 \]
 \[
 B_m(s) := 2(sa_{3m}^ra_{2m}^t - a_{2m}^ra_{3m}^t)
 \]
\[
C_m := (\tau_{\ell 0})^2 -G_{\ell m}-2a_{1m}^ra_{1m}^t,
\]
    
\begin{subequations} \label{eq:abc}
\begin{align}
    A_{\ell m}(s) &:= 2(a_{2m}^ra_{2m}^t + sa_{3m}^ra_{3m}^t)\\
    B_{\ell m}(s) &:= 2(sa_{3m}^ra_{2m}^t - a_{2m}^ra_{3m}^t)\\
    C_{\ell m} &:= (c\tau_{\ell m})^2 -G_{\ell m}-2a_{1m}^ra_{1m}^t,
\end{align}
\end{subequations}

and 

\[
    x_\ell = \cos \gamma_{\ell 0}, \quad
    y_\ell = \cos \gamma_{\ell 0}. 
\]

\begin{equation} \label{eq:xyfit}
    x_\ell = \cos \gamma^t_{\ell 0}, \quad
    y_\ell = \sin \gamma^t_{\ell 0}. 
\end{equation}

To find the solution to \eqref{eq:linabc},
we minimize 
\begin{equation} \label{eq:sxymin}
    (\widehat{s}_\ell,\widehat{x}_\ell,\widehat{y}_\ell)
    = \argmin_{s = \pm 1, x, y} J_{\ell}(s,x,y)
\end{equation}
where $J_{\ell}(\cdot)$ is the objective:
\begin{equation} \label{eq:Jfit}
    J_{\ell}(s,x,y) := \sum_{m=1}^M 
    \left( C_{\ell m} - A_{\ell m}(s)x - B_{\ell m}(s)y \right)^2.
\end{equation}
The minimization is \eqref{eq:sxymin} is easily performed:
For each value of $s= \pm 1$, the objective
\eqref{eq:Jfit} is a least squares with two unknowns.
Hence, the optimization \eqref{eq:sxymin}
has a unique minimum provided we have $M \geq 2$
measurements.
Once we obtain the minimum \eqref{eq:sxymin},
we obtain the parameters:
\begin{subequations} \label{eq:sgamfit}
\begin{align}
    s_{\ell} = s_{\ell 0} = \widehat{s}_\ell, \\
    \gamma_\ell = \arctan( \widehat{x}_\ell,\widehat{y}_\ell).
\end{align}
\end{subequations}

\noindent
\emph{Summary}:  
The procedure can be summarized as follows:
\begin{enumerate}
    \item Select $M+1$ TX-RX pairs $(\nbx^t_m,\nbx^r_m)$, 
    $m=0,\ldots,M$ with $M \geq 2$.
    \item Perform ray tracing to obtain the PWA parameters
    \eqref{eq:meas_rmdp} between each TX-RX pair.
    \item Sort the paths $\ell$ of the reference pair,
    $m=0$, in descending order of $|g_{\ell 0}|$.
    That is, sort the paths from strongest to weakest.
    \item Copy the PWA parameters at the reference
    pair to the RM model using \eqref{eq:pwa_rm}.
    This leaves only the parameters $s_\ell$ and $\gamma_\ell$ to be estimated.    
    \item For all displaced pairs, $m=1,\ldots,M$, perform the path matching and sort the paths in the order
    of matching with reference pair.
    \item For each path $\ell$ and displaced pair $m$, compute $\nba_{\ell m}^r$, $\nba_{\ell m}^t$ from
    \eqref{eq:adeffit}.  Also, compute 
    $G_{\ell m}$ in \eqref{eq:dsqfit3}.
    \item For each path $\ell$, compute the direction
    vectors $\nbu^r_\ell$ and $\nbu^t_\ell$
    from \eqref{eq:usph_fit}.
    \item For each $s=\pm 1$ and $\ell$ and $m$,
    compute $A_{\ell m}(s)$, $B_{\ell m}(s)$ and $C_{\ell m}$ in \eqref{eq:abc}.
    \item Perform the minimization in \eqref{eq:sxymin}
    to obtain $\widehat{s}_\ell,\widehat{x}_\ell,\widehat{y}_\ell$.
    The minimization is performed with two 
    linear least squares:  One with $s=1$ and the second
    with $s=-1$.

    \item Set $s_\ell$ and $\gamma_\ell$ from
    \eqref{eq:sgamfit}.

\end{enumerate}

}

\bibliographystyle{IEEEtran}
\bibliography{bibl}

\end{document}